\documentclass[twocolumn]{aastex63}

\shorttitle{MAPS IV: Vertical Disk Chemical Structures}
\shortauthors{Law et al.}

\usepackage[caption=false]{subfig}

\begin{document}

\title{Molecules with ALMA at Planet-forming Scales (MAPS) IV: Emission Surfaces and Vertical Distribution of Molecules}

\correspondingauthor{Charles J. Law}
\email{charles.law@cfa.harvard.edu}

\author[0000-0003-1413-1776]{Charles J. Law}
\affiliation{Center for Astrophysics \textbar\, Harvard \& Smithsonian, 60 Garden St., Cambridge, MA 02138, USA}

\author[0000-0003-1534-5186]{Richard Teague}
\affiliation{Center for Astrophysics \textbar\, Harvard \& Smithsonian, 60 Garden St., Cambridge, MA 02138, USA}

\author[0000-0002-8932-1219]{Ryan A. Loomis}
\affiliation{National Radio Astronomy Observatory, 520 Edgemont Rd., Charlottesville, VA 22903, USA}

\author[0000-0001-7258-770X]{Jaehan Bae}
\altaffiliation{NASA Hubble Fellowship Program Sagan Fellow}
\affil{Earth and Planets Laboratory, Carnegie Institution for Science, 5241 Broad Branch Road NW, Washington, DC 20015, USA}
\affiliation{Department of Astronomy, University of Florida, Gainesville, FL 32611, USA}

\author[0000-0001-8798-1347]{Karin I. \"Oberg}
\affiliation{Center for Astrophysics \textbar\, Harvard \& Smithsonian, 60 Garden St., Cambridge, MA 02138, USA}

\author[0000-0002-1483-8811]{Ian Czekala}
\altaffiliation{NASA Hubble Fellowship Program Sagan Fellow}
\affiliation{Department of Astronomy and Astrophysics, 525 Davey Laboratory, The Pennsylvania State University, University Park, PA 16802, USA}
\affiliation{Center for Exoplanets and Habitable Worlds, 525 Davey Laboratory, The Pennsylvania State University, University Park, PA 16802, USA}
\affiliation{Center for Astrostatistics, 525 Davey Laboratory, The Pennsylvania State University, University Park, PA 16802, USA}
\affiliation{Institute for Computational \& Data Sciences, The Pennsylvania State University, University Park, PA 16802, USA}
\affiliation{Department of Astronomy, 501 Campbell Hall, University of California, Berkeley, CA 94720-3411, USA}

\author[0000-0003-2253-2270]{Sean M. Andrews}
\affiliation{Center for Astrophysics \textbar\, Harvard \& Smithsonian, 60 Garden St., Cambridge, MA 02138, USA}

\author[0000-0003-3283-6884]{Yuri Aikawa}
\affiliation{Department of Astronomy, Graduate School of Science, The University of Tokyo, Tokyo 113-0033, Japan}

\author[0000-0002-2692-7862]{Felipe Alarc\'on}
\affiliation{Department of Astronomy, University of Michigan, 323 West Hall, 1085 South University Avenue, Ann Arbor, MI 48109, USA}

\author[0000-0003-4179-6394]{Edwin A.\ Bergin}
\affiliation{Department of Astronomy, University of Michigan, 323 West Hall, 1085 South University Avenue, Ann Arbor, MI 48109, USA}

\author[0000-0002-8716-0482]{Jennifer B. Bergner} 
\altaffiliation{NASA Hubble Fellowship Program Sagan Fellow}
\affiliation{University of Chicago Department of the Geophysical Sciences, Chicago, IL 60637, USA}

\author[0000-0003-2014-2121]{Alice S. Booth}
\affiliation{Leiden Observatory, Leiden University, 2300 RA Leiden, the Netherlands}
\affiliation{School of Physics and Astronomy, University of Leeds, Leeds, UK, LS2 9JT}

\author[0000-0003-4001-3589]{Arthur D. Bosman}
\affiliation{Department of Astronomy, University of Michigan, 323 West Hall, 1085 South University Avenue, Ann Arbor, MI 48109, USA}

\author[0000-0002-0150-0125]{Jenny K. Calahan} 
\affiliation{Department of Astronomy, University of Michigan, 323 West Hall, 1085 South University Avenue, Ann Arbor, MI 48109, USA}

\author[0000-0002-2700-9676]{Gianni Cataldi}
\affiliation{Department of Astronomy, Graduate School of Science, The University of Tokyo, Tokyo 113-0033, Japan}
\affiliation{National Astronomical Observatory of Japan, 2-21-1 Osawa, Mitaka, Tokyo 181-8588, Japan}

\author[0000-0003-2076-8001]{L. Ilsedore Cleeves}
\affiliation{Department of Astronomy, University of Virginia, Charlottesville, VA 22904, USA}

\author[0000-0002-2026-8157]{Kenji Furuya}
\affiliation{National Astronomical Observatory of Japan, 2-21-1 Osawa, Mitaka, Tokyo 181-8588, Japan}

\author[0000-0003-4784-3040]{Viviana V. Guzm\'{a}n}
\affiliation{Instituto de Astrof\'isica, Pontificia Universidad Cat\'olica de Chile, Av. Vicu\~na Mackenna 4860, 7820436 Macul, Santiago, Chile}

\author[0000-0001-6947-6072]{Jane Huang}
\altaffiliation{NASA Hubble Fellowship Program Sagan Fellow}
\affiliation{Center for Astrophysics \textbar\, Harvard \& Smithsonian, 60 Garden St., Cambridge, MA 02138, USA}
\affiliation{Department of Astronomy, University of Michigan, 323 West Hall, 1085 South University Avenue, Ann Arbor, MI 48109, USA}

\author[0000-0003-1008-1142]{John~D.~Ilee}
\affil{School of Physics and Astronomy, University of Leeds, Leeds, UK, LS2 9JT}

\author[0000-0003-1837-3772]{Romane Le Gal}
\affiliation{Center for Astrophysics \textbar\, Harvard \& Smithsonian, 60 Garden St., Cambridge, MA 02138, USA}
\affiliation{Univ. Grenoble Alpes, CNRS, IPAG, F-38000 Grenoble, France}
\affiliation{IRAP, Universit\'{e} de Toulouse, CNRS, CNES, UT3, 31400 Toulouse, France}
\affiliation{IRAM, 300 rue de la piscine, F-38406 Saint-Martin d'H\`{e}res, France}

\author[0000-0002-7616-666X]{Yao Liu}
\affiliation{Purple Mountain Observatory \& Key Laboratory for Radio Astronomy, Chinese Academy of Sciences, Nanjing 210023, China}

\author[0000-0002-7607-719X]{Feng Long}
\affiliation{Center for Astrophysics \textbar\, Harvard \& Smithsonian, 60 Garden St., Cambridge, MA 02138, USA}

\author[0000-0002-1637-7393]{Fran\c cois M\'enard}
\affiliation{Univ. Grenoble Alpes, CNRS, IPAG, F-38000 Grenoble, France}

\author[0000-0002-7058-7682]{Hideko Nomura}
\affiliation{National Astronomical Observatory of Japan, 2-21-1 Osawa, Mitaka, Tokyo 181-8588, Japan}

\author[0000-0002-1199-9564]{Laura M. P\'erez}
\affiliation{Departamento de Astronom\'ia, Universidad de Chile, Camino El Observatorio 1515, Las Condes, Santiago, Chile}

\author[0000-0001-8642-1786]{Chunhua Qi}
\affiliation{Center for Astrophysics \textbar\, Harvard \& Smithsonian, 60 Garden St., Cambridge, MA 02138, USA}

\author[0000-0002-6429-9457]{Kamber R. Schwarz}
\altaffiliation{NASA Hubble Fellowship Program Sagan Fellow}
\affiliation{Lunar and Planetary Laboratory, University of Arizona, 1629 E. University Blvd, Tucson, AZ 85721, USA}

\author[0000-0001-8455-7954]{Daniela Soto}
\affiliation{Instituto de Astrof\'isica, Pontificia Universidad Cat\'olica de Chile, Av. Vicu\~na Mackenna 4860, 7820436 Macul, Santiago, Chile}

\author[0000-0002-6034-2892]{Takashi Tsukagoshi}
\affiliation{National Astronomical Observatory of Japan, 2-21-1 Osawa, Mitaka, Tokyo 181-8588, Japan}

\author[0000-0003-4099-6941]{Yoshihide Yamato}
\affiliation{Department of Astronomy, Graduate School of Science, The University of Tokyo, Tokyo 113-0033, Japan}

\author[0000-0002-2555-9869]{Merel L. R. van 't Hoff}
\affiliation{Department of Astronomy, University of Michigan, 323 West Hall, 1085 South University Avenue, Ann Arbor, MI 48109, USA}

\author[0000-0001-6078-786X]{Catherine Walsh}
\affiliation{School of Physics and Astronomy, University of Leeds, Leeds, UK, LS2 9JT}

\author[0000-0003-1526-7587]{David J. Wilner}
\affiliation{Center for Astrophysics \textbar\, Harvard \& Smithsonian, 60 Garden St., Cambridge, MA 02138, USA}

\author[0000-0002-0661-7517]{Ke Zhang}
\altaffiliation{NASA Hubble Fellow}
\affiliation{Department of Astronomy, University of Michigan, 323 West Hall, 1085 South University Avenue, Ann Arbor, MI 48109, USA}
\affiliation{Department of Astronomy, University of Wisconsin-Madison, 475 N Charter St, Madison, WI 53706}



\begin{abstract}
The Molecules with ALMA at Planet-forming Scales (MAPS) Large Program provides a unique opportunity to study the vertical distribution of gas, chemistry, and temperature in the protoplanetary disks around IM~Lup, GM~Aur, AS~209, HD~163296, and MWC~480. By using the asymmetry of molecular line emission relative to the disk major axis, we infer the emission height~($z$) above the midplane as a function of radius~($r$). Using this method, we measure emitting surfaces for a suite of CO isotopologues, HCN, and C$_2$H. We find that $^{12}$CO emission traces the most elevated regions with $z/r>0.3$, while emission from the less abundant $^{13}$CO and C$^{18}$O probes deeper into the disk at altitudes of $z/r~\lesssim~0.2$. C$_2$H and HCN have lower opacities and SNRs, making surface fitting more difficult, and could only be reliably constrained in AS~209, HD~163296, and MWC~480, with $z/r~\lesssim~0.1$, i.e., relatively close to the planet-forming midplanes. We determine peak brightness temperatures of the optically thick CO isotopologues and use these to trace 2D disk temperature structures. Several CO temperature profiles and emission surfaces show dips in temperature or vertical height, some of which are associated with gaps and rings in line and/or continuum emission. These substructures may be due to local changes in CO column density, gas surface density, or gas temperatures, and detailed thermo-chemical models are necessary to better constrain their origins and relate the chemical compositions of elevated disk layers with those of planet-forming material in disk midplanes. This paper is part of the MAPS special issue of the Astrophysical Journal Supplement.
\end{abstract}



\section{Introduction} \label{sec:intro}

Protoplanetary disks are highly stratified in their physical and chemical properties \citep[][]{Williams11} with flared emitting surfaces set by the balance of hydrostatic equilibrium, as first recognized in their spectral energy distributions \citep{Kenyon87}. In particular, vertical gradients in gas temperature, density, radiation, and ionization result in a rich chemical structure over the height of the disk \citep[e.g.,][]{vanZadelhoff01, Woitke09, Walsh10, Fogel11}. The efficiency of vertical mixing \citep{Ilgner04, Semenov11} and the presence of meridional flows driven by embedded planets \citep{Morbidelli14, Dong19, Teague19Natur} also influence the vertical distribution of molecular material in disks. Vertical chemical structures have been seen in observations of highly inclined disks, which allow the emission distribution to be mapped directly \citep{Dutrey17, Teague20_goham, Podio20, RR21}. In more moderately inclined disks, the excitation temperatures of different species and molecular isotopologues have instead been used to infer the properties of the vertical gas distribution \citep{Dartois_ea_2003, Pietu07, Oberg20_Rosetta, Cleeves21}. A detailed understanding of this vertical structure is required to interpret these observations and to assess how well connected the molecular gas abundances derived from line observations are to those of the planet-forming material in disk midplanes.

The high spatial and spectral resolutions offered by ALMA allow for the direct measurement of the height at which molecular emission arises for mid-inclination disks. With sufficient angular resolution and surface brightness sensitivity, it is possible to spatially resolve emission arising from elevated regions above and below the midplane \citep[e.g.,][]{deGregorio-Monsalvo_ea_2013, Rosenfeld13, Isella18, Huang20_GMAur}. In cases such as these, the emission surface of molecular lines can be directly extracted with a technique similar to that used in NIR observations to infer scattering surfaces \citep[e.g.,][]{Monnier17, Avenhaus18}. Such a method was first presented by \citet{Pinte18}, who used it to map the CO, $^{13}$CO, and C$^{18}$O 2--1 emission surfaces in IM~Lup. As the less abundant isotopologues probe deeper in the disk (i.e., closer to the midplane), this also allows for an empirical derivation of the two dimensional gas temperature structure, an essential input for models and simulations of planet formation. A similar approach has now been employed to map CO isotopologue surfaces in a handful of disks \citep[e.g.,][]{Teague19Natur, Paneque21, Rich21}.

As part of the Molecules with ALMA at Planet-forming Scales (MAPS) Large Program, five protoplanetary disks were observed in several molecular lines expected to emit strongly at different vertical locations. In this paper, we provide a framework for generalizing the method presented in \citet{Pinte18} to a larger sample of disks. We use this framework to characterize line emission heights, gas temperatures, and disk vertical substructures. The layout of the paper is as follows: we present a brief overview of the observations in Section \ref{sec:obs}. In Section \ref{sec:deriving_emission_surfaces}, we describe how emission surfaces were derived and fit with analytical functions. We calculate the radial and vertical temperature profiles and compare the observed vertical structures with previous millimeter and NIR observations in Section \ref{sec:results}. In Section \ref{sec:discussion}, we discuss possible origins of disk vertical substructures and summarize our findings in Section \ref{sec:conclusions}. All publicly-available data products are listed in Section \ref{sec:VADPs}.

\section{Observations} \label{sec:obs}

The observations used in this study were obtained as part of the MAPS ALMA Large Program (2018.1.01055.L), which targeted the protoplantary disks around IM~Lup, GM~Aur, AS~209, HD~163296, and MWC~480. An overview of the survey, including observational setup, calibration, and rationale, is provided in \citet{oberg20}, while the imaging process is described in detail in \citet{czekala20}. The analysis in this work is based primarily on the ALMA Band 6 images generated with a robust parameter of 0.5, rather than the fiducial images presented in the overview paper. We opted to use these images to leverage their higher spatial resolutions (10--40\% smaller total beam area) relative to the fiducial images with a 0\farcs15 circularized beam.  The major and minor axes of the synthesized beam ranged from 0\farcs13--0\farcs17 and 0\farcs10--0\farcs13, respectively. At the distances of the MAPS disks, these correspond to physical scales of ${\sim}$14~au--27~au and 10~au--20~au, respectively. For extracting brightness temperatures, we also made use of the corresponding non-continuum-subtracted image cubes, which were imaged in the same way as the line-only data.

This work is based on the following five species: CO, $^{13}$CO, C$^{18}$O, HCN, and C$_2$H. The primary focus for the CO isotopologues is on the Band 6 transitions, i.e., J=2--1, as they possess the highest spatial resolutions. For HCN and C$_2$H, we only considered the brightest hyperfine components of the Band 6 transitions, i.e., C$_2$H N=3--2, J=$\frac{7}{2}$--$\frac{5}{2}$, F=4--3 and HCN J=3--2, F=3--2. For simplicity, we refer to these lines as HCN 3--2 and C$_2$H 3--2, respectively. This set of molecules and lines was selected for this analysis as they are consistently the brightest in the MAPS sample and possessed radially extended emission that allowed for the determination of robust emission surfaces \citep{law20b}. 

Since the $^{13}$CO and C$^{18}$O isotopologues were observed in both Bands 3 and 6, we also aimed to assess the influence of excitation on the derived surfaces, namely if different transitions from the same molecule emit at different disk heights. To do so, we used the tapered (0\farcs30) images \citep[see Section 6.2,][]{czekala20}, which allowed us to match the spatial resolutions of the 2--1 (Band 6) and 1--0 (Band 3) lines for both species. This comparison, however, did not yield any conclusive results (see Appendix \ref{sec:app:Band3_excitation}).

\section{Emission Surfaces} \label{sec:deriving_emission_surfaces}

In the following subsections, we present an outline of how we derived emitting layers starting from the line image cubes to the final data products. We then describe how we fit an analytical function to each of the surfaces.

\subsection{Deriving Emitting Layers} \label{sec:fitting_surfaces}

\begin{figure*}[ht]
\centering
\includegraphics[width=\linewidth]{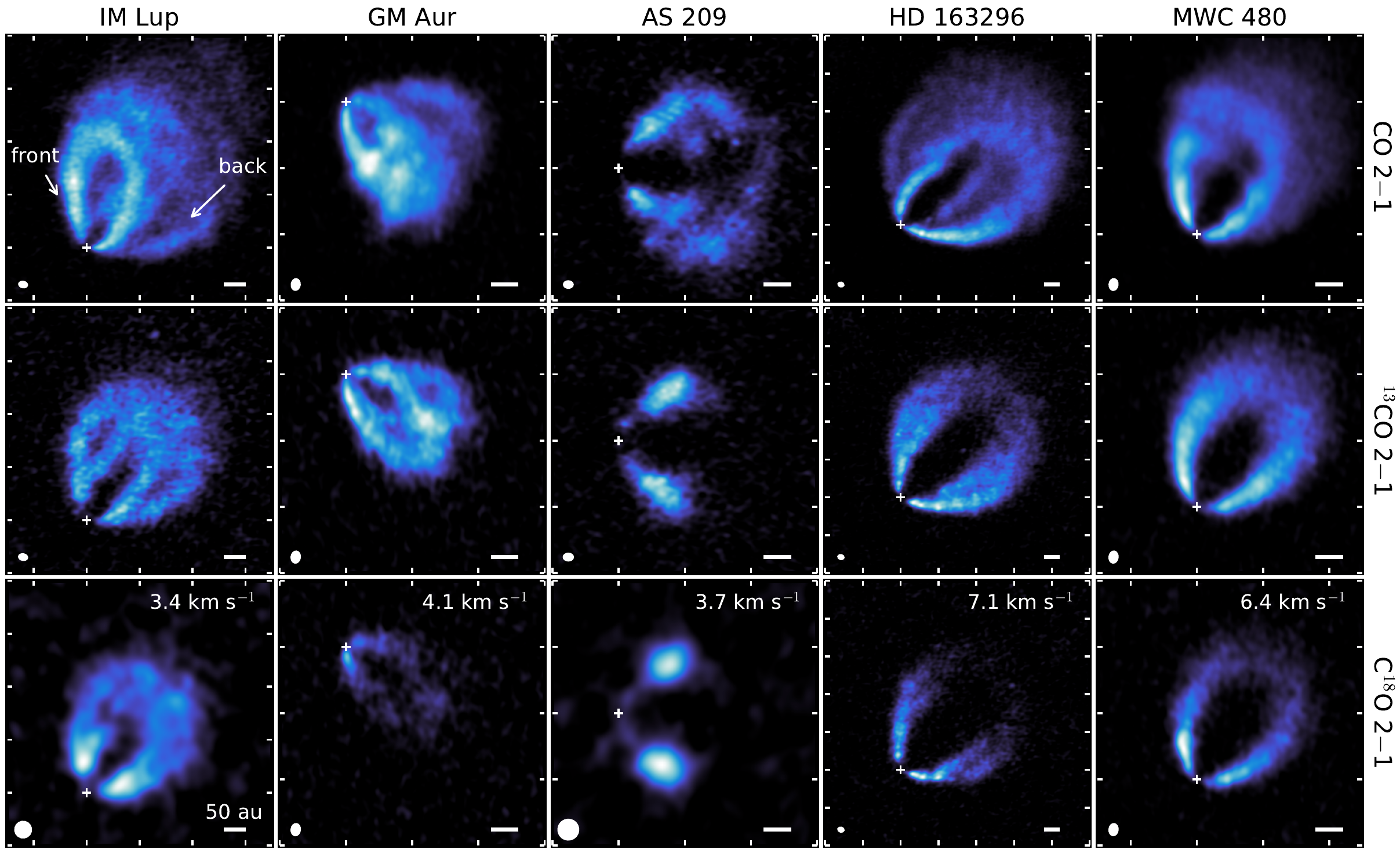}
\caption{Representative channels for the MAPS sample, ordered from left to right by increasing stellar mass (see Table 1 in \citet{oberg20}), for CO 2--1, $^{13}$CO 2--1, and C$^{18}$O 2--1. Both sides of the disk are visible and can be used to fit for the emitting layer. The C$^{18}$O 2--1 images in IM~Lup and GM~Aur have been tapered to 0\farcs30, as described in Section \ref{sec:overview_of_emission_surfaces}. LSRK velocities are shown in the upper right corner and the plus symbol indicates the disk center. Axes are angular offsets from the disk center with 1$^{\prime \prime}$ tick marks. An arcsinh color stretch has been applied to highlight faint outer structures.}
\label{fig:Example_Channel_Disk_Sides}
\end{figure*}

The emission surfaces derived in this work represent the mean height of the emission surface for each molecular tracer, or put simply, where the bulk of the emission arises from in each line and disk. We extracted these emission heights from the image cubes by using the asymmetry of the emission relative to the major axis of the disk. By assuming that disks are azimuthally symmetric and that the gas is on circular orbits, this allows us to infer the height of the emission above the disk midplane. A prerequisite for this approach is the ability to spatially resolve the front and back of the disk in multiple channels, as illustrated for CO 2--1, $^{13}$CO 2--1, and C$^{18}$O 2--1 in Figure \ref{fig:Example_Channel_Disk_Sides}. For each line and disk, we determined if both disk sides were sufficiently spatially-resolved via visual inspection and then confirmed that the predicted isovelocity contours matched the spatial distribution of line emission in the channel maps (see Appendix \ref{sec:app:isovelocity}). We refer readers to \citet{Pinte18} for additional details about this method.

We derived emission surfaces using the \texttt{disksurf}\footnote{\url{https://github.com/richteague/disksurf}} Python package, which implements this method while providing additional functionality to filter the data to extract more precise emission surfaces. This series of filtering steps is described in detail in Appendix \ref{sec:app:surface_extraction}. We used the \texttt{get\_emission\_surface} function to extract the deprojected radius $r$, emission height $z$, surface brightness I$_{\nu}$, and channel velocity $v$ for each pixel associated with the emitting surface. These surfaces represent individual measurements, i.e., pixels, from the line image cubes.

\begin{figure}[ht]
\centering
\includegraphics[width=\linewidth]{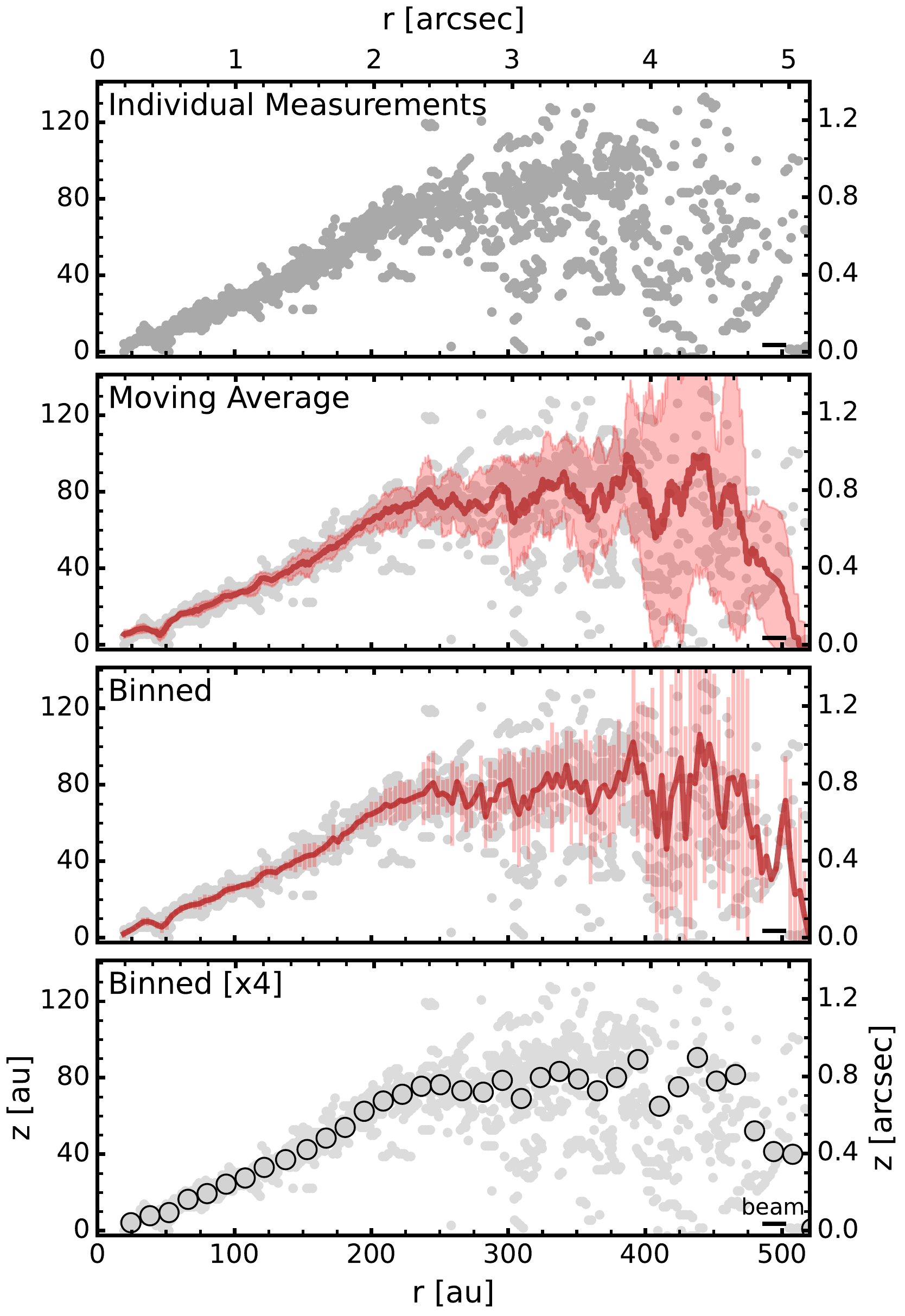}
\caption{Comparison of emission heights derived for CO 2$-1$ in HD~163296 for individual measurements, moving average, and radially-binned surfaces (from top to bottom). For increased visual clarity, we occasionally bin surfaces by an extra factor of a few, as illustrated in the bottom panel. The FWHM of the major axis of the synthesized beam is shown in the bottom right of each panel.}
\label{fig:Example_Surface_Dataproducts}
\end{figure}

We then use two different methods to further reduce scatter in the individual emission surface measurements and help better identify substructure. First, we radially bin the individual measurements using bin sizes of 1/4 of the FWHM of the beam major axis, i.e., the same as the radial intensity profiles in \citet{law20b}. The uncertainty is given by the standard deviation in each bin. We note that \citet{Pinte18} included the uncertainty in the disk inclination in these uncertainties. We opted not to do this as the disk inclination is a systematic uncertainty and results in a scaling of the vertical height axis, and not a relative uncertainty between radial bins. Besides binning, we also calculated the moving average and standard deviation of the individual surface measurements. As the spacing between radial points is not uniform, we used a window size with a minimum size of one quarter of the FWHM of the beam major axis. This window was required to contain a fixed number of points, which means that the physical size that it represents changes with radius due to the non-uniform radial sampling from the deprojection process, i.e., in the less dense, outer disk, the window expands in order to still encompass this fixed number of points. A summary of these different data products is shown in Figure \ref{fig:Example_Surface_Dataproducts}. While it was found that binned and moving average surfaces showed the same trends, the binned surface benefited from a uniform radial sampling, while the moving average retained a finer radial sampling, essential for identifying subtle perturbations associated with features in the dust continuum.

All three types of line emission surfaces --- individual measurements, radially-binned, and moving averages --- are provided as Value-Added Data Product (VADPs) and will be made available to the community through our dedicated website hosted by ALMA (\url{https://almascience.nrao.edu/alma-data/lp/maps}). See Section \ref{sec:VADPs} for further details. Throughout this work, we sometimes bin these data products further for visual clarity, but all quantitative analysis is done using the original binning of each type of emission surface.

\subsection{Analytical Fits} \label{sec:surface_fitting}

To facilitate implementing these emission surfaces in models and for comparison with other observations, we fit an exponentially-tapered power law to all CO emission surfaces. This parametric fit was chosen as it describes the flared surface in the inner disk ($\lesssim$~200~au) and captures the expected drop in the outer disk due to decreasing gas surface density, as seen in Figure \ref{fig:Example_Surface_Dataproducts}. We adopt the following functional form:

\begin{equation} \label{eqn:exp_taper}
z(r) = z_0 \times \left( \frac{r}{1^{\prime \prime}} \right)^{\phi} \times \exp \left(- \left[ \frac{r}{r_{\rm{taper}}} \right]^{\psi} \right)
\end{equation}

\noindent where $z_0$, $\phi$, and $\psi$ should always be non-negative. A value of $\phi > 1$ indicates that $z$ increases with radius, while $0 < \phi < 1$ tends toward a flat $z(r)$ profile. When $r_{\rm{taper}} \gg 1^{\prime \prime}$, $z_0$ represents the $z/r$ value at 1$^{\prime \prime}$. Note that some previous works, e.g., \citet{Teague19Natur}, instead used a double power law profile to capture the drop in emission height at large radius. It was found that this tapered form, on average, provided a better fit to the data with less manual tuning required.

To ensure the robustness of these fits, we also restricted the radial range used for fitting to locations with high densities of $(r,\,z)$ measurements. The radial ranges used in each fit are given by $r_{\rm{fit, in}}$ and $r_{\rm{fit, out}}$ in Table \ref{tab:emission_surf}.

\begin{deluxetable*}{llccccccc}
\tablecaption{Parameters for Emission Surface Fits\label{tab:emission_surf}}
\tablewidth{0pt}
\tablehead{
\colhead{Source} & \colhead{Line} & \colhead{Velocity Range} & \multicolumn6c{Exponentially-Tapered Power Law$^a$} \\ \cline{4-9}
\colhead{} & \colhead{} & \colhead{[km~s$^{-1}$]} & \colhead{$r_{\rm{fit, in}}$ [$^{\prime \prime}$]} & \colhead{$r_{\rm{fit, out}}$ [$^{\prime \prime}$]} &\colhead{$z_0$ [$^{\prime \prime}$]} & \colhead{$\phi$} &\colhead{$r_{\rm{taper}}$ [$^{\prime \prime}$]} &\colhead{$\psi$}}
\startdata
IM Lup & CO 2$-$1 & [2.8, 6.4] & 0.21 & 3.26 & 4.37 & 3.144 & 0.254 & 0.655\\
 & $^{13}$CO 2$-$1 & [2.6, 6.4] & 0.61 & 2.02 & 0.159 & 2.599 & 1.928 & 4.993\\
GM Aur & CO 2$-$1 & [3.1, 7.5] & 0.08 & 3.36 & 0.385 & 1.066 & 3.767 & 4.988\\
 & $^{13}$CO 2$-$1 & [3.7, 7.1] & 0.2 & 1.89 & 0.113 & 4.539 & 1.496 & 4.989\\
 & C$^{18}$O 2$-$1 & [$-$2.9, 13.9] & 0.21 & 0.66 & 0.95 & 3.556 & 0.402 & 3.766\\
AS 209 & CO 2$-$1 & [2.9, 6.5] & 0.07 & 1.98 & 0.219 & 1.292 & 1.786 & 4.854\\
 & $^{13}$CO 2$-$1 & [2.9, 6.5] & 0.73 & 1.35 & 0.175 & 2.98 & 1.124 & 2.445\\
HD 163296 & CO 2$-$1 & [4.3, 13.5] & 0.19 & 4.73 & 0.388 & 1.851 & 2.362 & 1.182\\
 & $^{13}$CO 2$-$1 & [3.5, 13.5] & 0.31 & 3.42 & 0.121 & 1.503 & 3.158 & 4.996\\
 & C$^{18}$O 2$-$1 & [3.5, 8.1] & 0.39 & 1.43 & 0.174 & 2.956 & 1.043 & 4.994\\
MWC 480 & CO 2$-$1 & [2.8, 7.4] & 0.13 & 3.69 & 0.261 & 1.35 & 3.098 & 3.074\\
 & $^{13}$CO 2$-$1 & [2.8, 7.2] & 0.05 & 2.26 & 1.248 & 2.165 & 0.215 & 0.683\\
 & C$^{18}$O 2$-$1 & [$-$8.2, 18.2] & 0.13 & 1.46 & 0.065 & 1.37 & 0.961 & 4.834\\
\enddata
\tablenotetext{a}{The statistical uncertainties on the fitted parameters were typically $\lesssim1\%$, but this does not account for the systematic uncertainties associated with extracting the emission surfaces, which, although more difficult to quantify, are likely substantially larger than those related to the analytical fits.}
\end{deluxetable*}

We used the Affine-invariant MCMC sampler \citep{Goodman10} implemented in \texttt{emcee} \citep{Foreman_Mackey13} to estimate the posterior distributions of these fits. We used 64 walkers which take 1000 steps to burn in and an additional 500 steps to sample the posterior distribution function. We chose an MCMC fitting approach rather than a simple chi-squared minimization to better handle the degeneracies between fitted parameters. Individual pixels are not all necessarily independent as they may originate within a single beam, which can lead to an underestimation of the true uncertainties on the extracted heights, i.e., on how well we can extract the mean $z$ from a sample of columns. However, this will not necessarily affect the mean height itself. This is analogous to drawing random samples from a normal distribution, where given a sufficiently large number of samples, the standard deviation of those samples provides a good estimate of the uncertainty on the mean of the distribution. Instead, if these samples are correlated and, e.g., that for every second draw, the sample is biased towards the one immediately preceding it, we will over-sample the central region compared to the wings. In this case, the standard deviation of the total ensemble will underestimate the true standard deviation of the distribution, but will not alter the underlying mean value.

The presence of this potential spatial correlation between pixels does not affect the analytical fits, which are instead dominated by overall radial trends rather than the vertical scatter in height measurements. Fits were also performed using the individual measurements, binned, and moving average surfaces, and we confirmed that all produced consistent results. We found that fits to the moving average surfaces were slightly more reliable at the outer disk radii. This is likely because the moving average is a compromise in terms of the number of radial points and relative signal-to-noise ratio (SNR), as the raw data have a finer grid of radial points but much larger scatter, while the binned data have coarser radial information but higher SNR. 

The fitted $\phi$ (${\sim}$1.5-4) for the CO emission surfaces are often several times larger than the flaring indices of the gas pressure scale heights (1.08-1.35), which are derived by fitting the observed spectral energy distributions of each star+disk system \citep[for more details, see][]{zhang20}. This difference was previously noted by \citet{Pinte18} in IM~Lup, who ascribed it to the sharp drop-off in UV radiation from the central star. The stellar irradiation determines the shape of the emitting surface, which follows a layer of approximately constant optical depth from the perspective of the star, rather than tracing the disk scale height. In particular, in the inner ${<}$300~au (${<}$120~au for AS~209) where the surfaces are still steeply rising, the CO surface originates from a height that is 2.5-3.5 times the scale height, while the $^{13}$CO and C$^{18}$O surfaces are at approximately 1-1.5 scale heights. Similar values have been reported previously for IM~Lup \citep{Pinte18}, as well as in DM~Tau \citep{Dartois_ea_2003} and the Flying Saucer \citep{Dutrey17}.

\begin{figure*}[ht]
\centering
\includegraphics[width=\linewidth]{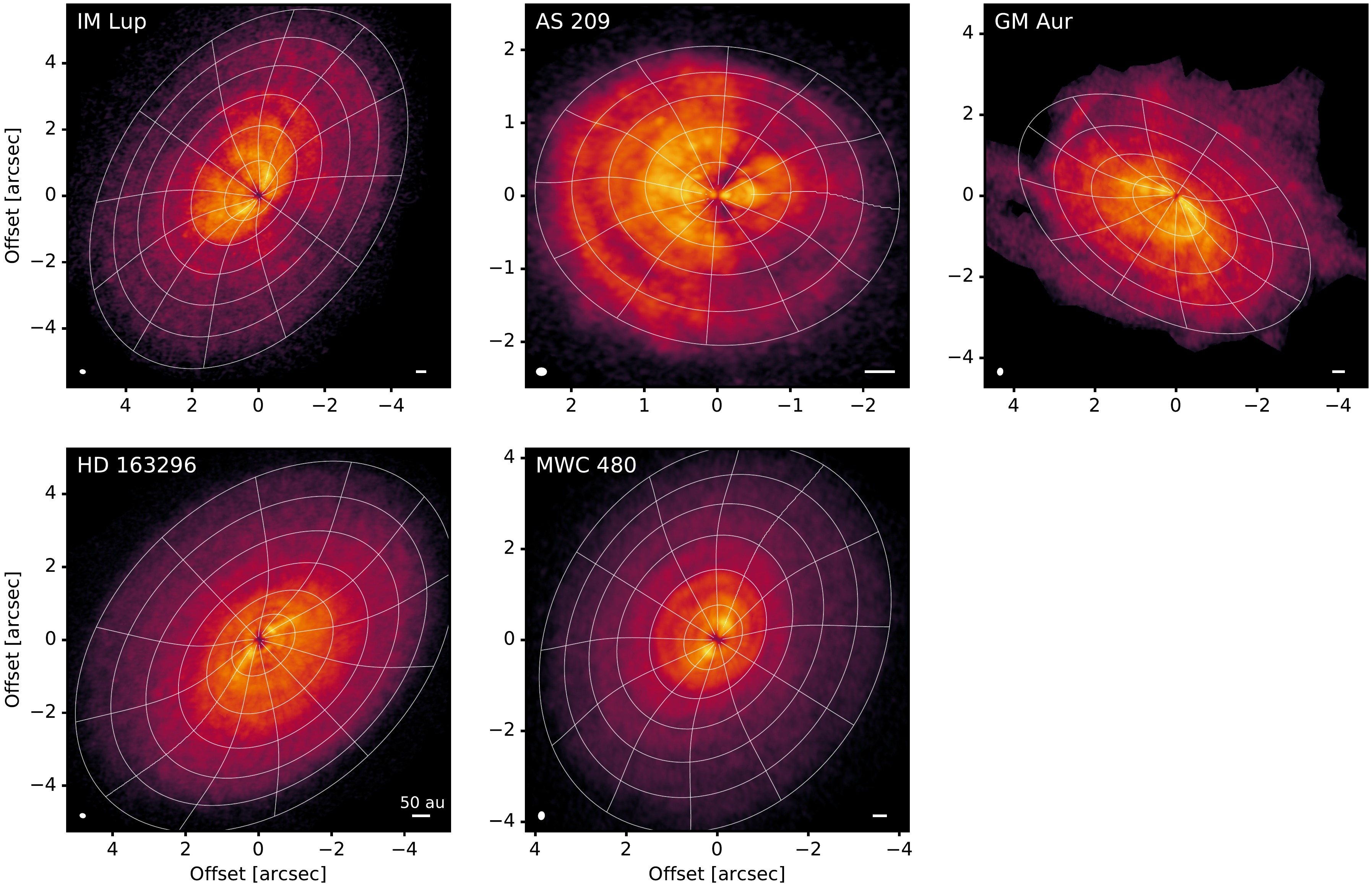}
\caption{Peak intensity maps of CO 2--1 for all MAPS sources with overlaid contours showing the fitting emission surfaces, as listed in Table \ref{tab:emission_surf}. The synthesized beam and a scale bar indicating 50~au is shown in the lower left and right corner, respectively, of each panel.}
\label{fig:Surface_Moment_Overlay}
\end{figure*}

To further illustrate the geometry of the fitted surfaces, Figure \ref{fig:Surface_Moment_Overlay} shows an overlay of the inferred emission surfaces on the peak intensity maps of CO 2--1 for all disks. Isovelocity contours generated using the surface fits from Table \ref{tab:emission_surf} for the CO isotopologues are also provided in Appendix \ref{sec:app:isovelocity}.

\section{Results} \label{sec:results}

\subsection{Overview of Emission Surfaces} \label{sec:overview_of_emission_surfaces}

\begin{figure*}[ht]
\centering
\includegraphics[width=\linewidth]{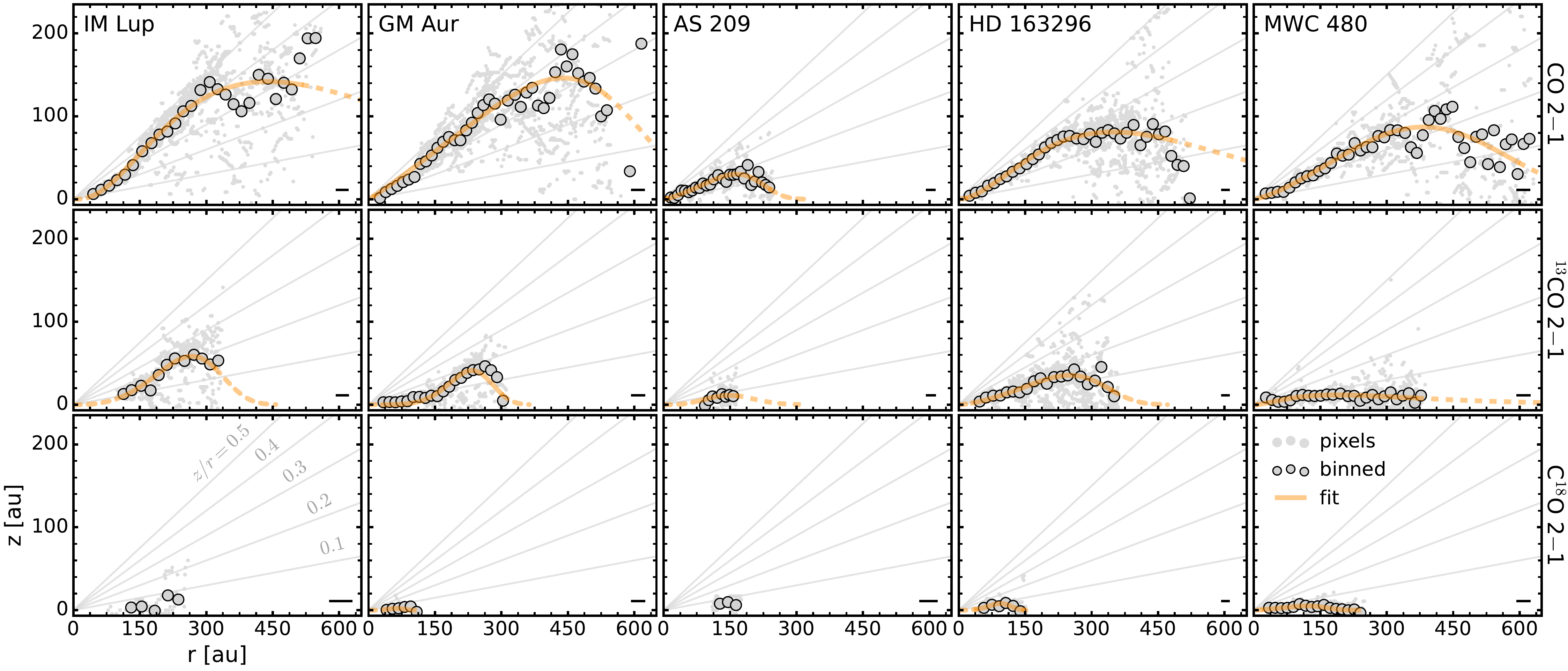}
\caption{Emission surfaces for CO 2--1, $^{13}$CO 2--1, and C$^{18}$O 2--1 in all MAPS sources. Large gray points show radially-binned surfaces and small, light gray points represent individual measurements. The orange lines show the exponentially-tapered power law fits from Table \ref{tab:emission_surf}. The solid lines show the radial range used in the fitting, while the dashed lines are extrapolations. The C$^{18}$O 2--1 surfaces in IM~Lup and AS~209 are tentative and were derived from the tapered (0\farcs30) images. Lines of constant $z/r$ are shown in gray. The FWHM of the major axis of the synthesized beam is shown in the bottom right corner of each panel.}
\label{fig:CO_B6_surfaces}
\end{figure*}

Figure \ref{fig:CO_B6_surfaces} shows the surfaces derived for CO isotopologues in all disks. For each disk, the CO 2--1 surface lies higher than that of the $^{13}$CO 2--1, which, in turn, is higher than C$^{18}$O 2--1. Such a progression is consistent with a line optical depth of ${\sim}$1 being reached at deeper layers for rarer isotopologues. There are considerable differences in the absolute surface heights of CO and $^{13}$CO between disks. For instance, the CO surface reaches a peak height of $z \approx$~200~au in IM~Lup and GM~Aur, while they are below ${\sim}$100~au in AS~209, HD~163296, and MWC~480. A similar trend is seen in $^{13}$CO, where IM~Lup has a maximum height of $z\approx$~100~au, while AS~209 and MWC~480 peak at $z<$~40~au. This range in absolute emission heights translates into a range of peak $z/r$. CO emission is present at $z/r \gtrsim 0.5$ in IM~Lup and GM~Aur, while $z/r \sim$~0.2 in AS~209. $^{13}$CO shows less overall variation between disks than CO, and is generally present at $z/r < 0.2$. C$^{18}$O has $z/r \lesssim 0.1$ towards those disks where we had enough signal to estimate emission heights. Finally, the relationship among the CO, $^{13}$CO, and C$^{18}$O emission heights within disks vary across the sample. In MWC~480, the CO emission surface is relatively elevated with $z/r \sim 0.3$, while $^{13}$CO and C$^{18}$O are both very flat, i.e., $z/r < 0.1$. By contrast, HD~163296 shows a gradual progression of $z/r\sim0.3$ to $0.2$ to 0.1 for CO, $^{13}$CO and C$^{18}$O, respectively. 

No surfaces could be derived from the full resolution images of C$^{18}$O 2--1 in IM~Lup and AS~209 due to insufficient SNR. However, we were able to extract surfaces from the corresponding tapered (0\farcs30) image cubes \citep[see Section 6.2,][]{czekala20} but consider these to be tentative and did not attempt to fit analytical functions to these surfaces. Both are shown in Figure \ref{fig:CO_B6_surfaces}, but are otherwise omitted from subsequent analysis.

In all CO lines, except for CO in IM~Lup, we see an initial increase of z/r with radius, i.e. flaring, a flattening, and then eventual turnover due to decreasing gas surface densities at large radii. IM~Lup, which is known to possess extended diffuse CO emission \citep{Cleeves16}, does not show clear evidence of this turnover and only shows moderate indications of flattening. All surfaces show some degree of vertical scatter, which is a combination of thermal noise in the images and potential azimuthal variations in the underlying emission surfaces. The relative contribution is, however, specific to each disk and emission line. As this scatter increases substantially with radius, the line SNR is likely the most important factor in setting the vertical scatter, at least at large radii. Due to different projections at varying azimuths in specific channels, the height of a particular pixel can often be easier or more difficult to determine, which provides an additional source of uncertainty in vertical pixel positions. For instance, channels with less favorable viewing geometries where the two disk sides cannot be easily distinguished make it harder to measure emission surface heights. This often occurs at velocities either very close to or substantially offset from the source systemic velocity. For example, in Figure \ref{fig:Isovelocity_IM_Lup}, channels at larger velocity offsets (${\leq}$2.6~km~s$^{-1}$ or ${\ge}$6.2~km~s$^{-1}$) and those near the systemic velocity (${\approx}$ 4.5~km~s$^{-1}$) show poorly-separated upper and lower disk surfaces.

Nonetheless, some surfaces appear more tightly constrained than others, i.e., CO in HD~163296 shows considerably less vertical dispersion compared to that of the MWC~480 disk. This scatter in the MWC~480 disk is not just due to noise, but is the result of localized azimuthal deviations. Perturbations in azimuthal velocity, on the order of a few \%, are located at ${\sim}$240, 340, 370, and 450~au in MWC~480 \citep{teague20}, which approximately align with regions of prominent vertical scatter in its CO emission surface. Similarly detailed and disk-specific analyses are required to discern the origins of vertical scatter in the other MAPS sources. Several disks also show evidence of substructure in their surfaces, e.g., C$^{18}$O in HD~163296, which we discuss in detail in Section \ref{sec:substructures_emisssion_surfaces}.

\begin{figure*}[ht]
\centering
\includegraphics[width=0.7\linewidth]{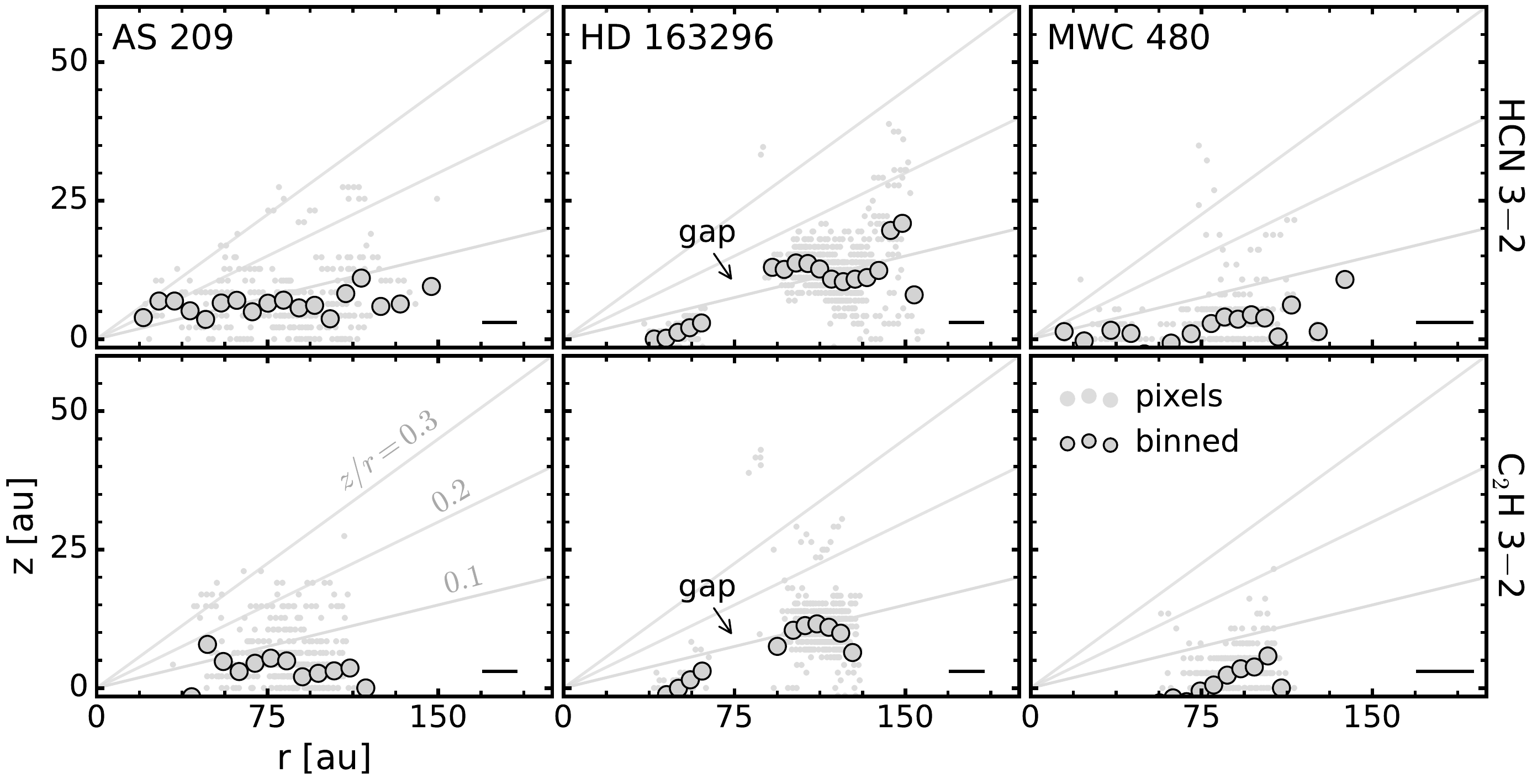} 
\caption{Emission surfaces for HCN 3--2 and C$_2$H 3--2 in AS~209, HD~163296, and MWC~480. Large gray points show radially-binned surfaces and small, light gray points represent individual measurements. Lines of constant $z/r$ are shown in gray. The FWHM of the major axis of the synthesized beam is shown in the bottom right corner of each panel. The HCN and C$_2$H surfaces for the remaining MAPS sources IM~Lup and GM~Aur lacked sufficient SNR for robust surface determinations.}
\label{fig:HCN_C2H_B6_surfaces}
\end{figure*}

Figure \ref{fig:HCN_C2H_B6_surfaces} shows the surfaces for C$_2$H and HCN. Of the disks around T Tauri stars, only surfaces for AS~209 could be extracted and appear to be at $z/r \lesssim 0.1$. The C$_2$H and HCN surfaces in IM~Lup and GM~Aur could not be reliably constrained due to their low line optical depths and SNRs compared to CO and $^{13}$CO. The two disks around Herbig Ae stars, HD~163296 and MWC~480, also show emission at a $z/r$ of 0.1 or less. In MWC~480, both HCN and C$_2$H are present at $z/r < 0.1$, similar to the $^{13}$CO and C$^{18}$O surfaces. HD~163296 is the only source where the C$_2$H and HCN lines show any structure; there is a clear gap in the surfaces corresponding to the gap between the two innermost rings in the radial emission profiles \citep{law20b}. The first ring at 45~au is less vertically extended with $z/r < 0.1$, while the emission in the second ring at 110~au is more elevated at $z/r \approx 0.1$. We do not attempt parametric fits for any of the HCN and C$_2$H lines.

\subsection{Comparison with NIR rings}

\begin{figure}[ht]
\centering
\includegraphics[width=\linewidth]{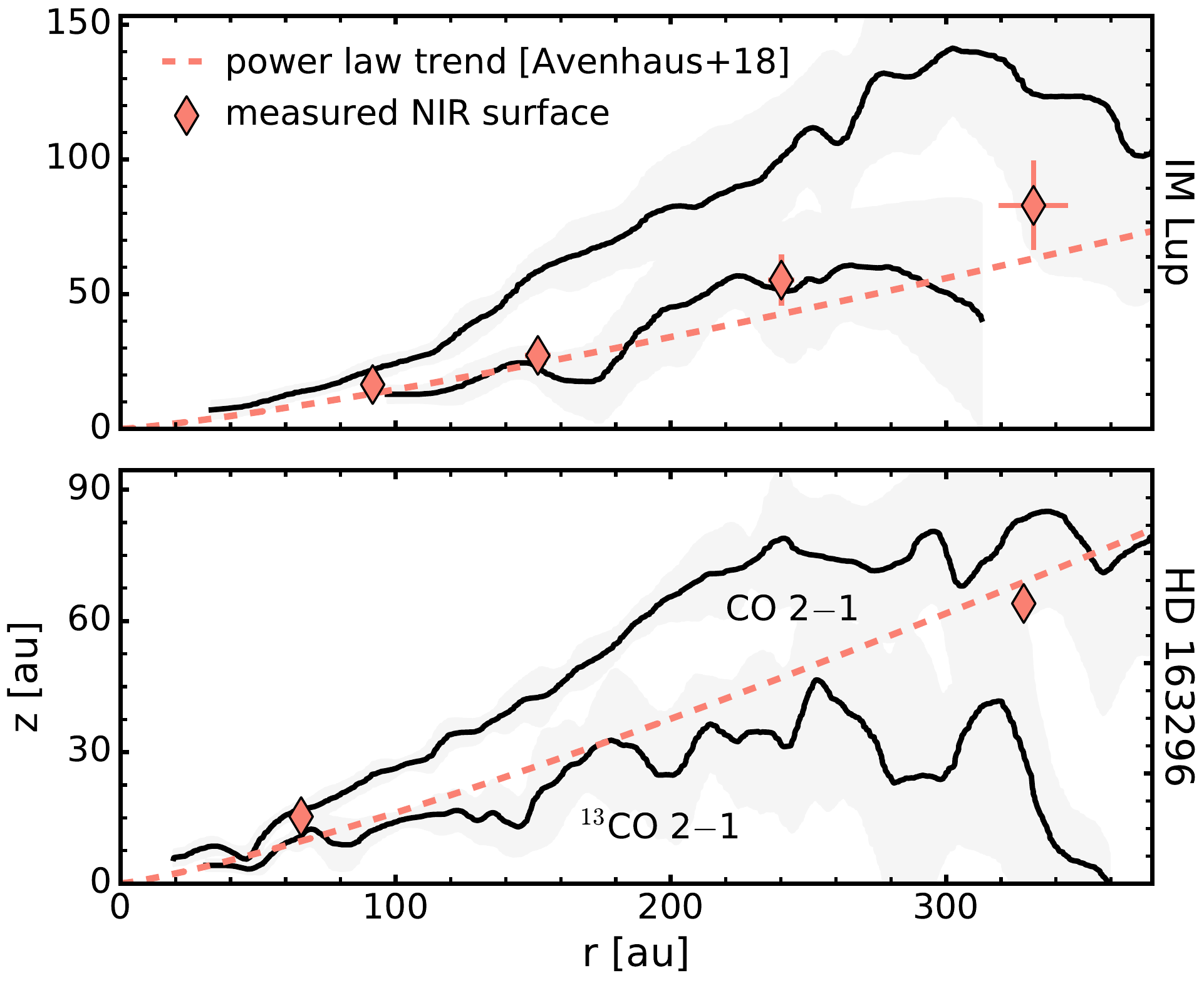}
\caption{Emission surfaces of CO and $^{13}$CO 2--1 in IM~Lup (top) and HD~163296 (bottom). The black lines are the moving average surfaces and gray shaded regions show the 1$\sigma$ uncertainty.  The red diamonds show individual height measurements of NIR rings \citep{Monnier17, Avenhaus18, Rich20}, while the red dashed line shows the inferred NIR surface using the relation found in a sample of disks in \citet{Avenhaus18}. The values from \citet{Monnier17} have been scaled to reflect the updated \textit{Gaia} distance for HD~163296. The errorbars are smaller than the marker for the innermost rings in both IM~Lup and HD~163296 and the 330~au ring in HD~163296 does not have reported uncertainties.}
\label{fig:CO_13CO_vs_NIR_Surface}
\end{figure}

All of the MAPS sources have been observed in scattered light \citep{Schneider03, Kusakabe12, Monnier17, Avenhaus18,Muro_Arena18}, which provides valuable information about the micron-sized dust grains in these disks. The IM~Lup, AS~209\footnote{If deprojected with a nonzero flaring angle, \citet{Avenhaus18} found that AS~209 possesses either one (112~au) or three (78, 140, and 243~au) NIR rings depending on whether the northern side is the near or far side, respectively. Subsequent observations \citep{Guzman18, Teague18_AS209} showed that the latter interpretation is correct.}, and HD~163296 disks have well-defined rings in the NIR, but only IM~Lup and HD~163296 have direct estimates of their NIR emitting surfaces, as measured from individual rings. The inner NIR ring in HD~163296 has a height measured from \citet{Monnier17}, while the outer ring at 330~au was recently found to have a dust scale height of 64~au in NIR/HST observations \citep{Rich20}. All four rings in IM~Lup have measured NIR heights \citep{Avenhaus18}. Figure \ref{fig:CO_13CO_vs_NIR_Surface} shows these NIR heights compared to the CO and $^{13}$CO 2--1 emission surfaces. We also plot the NIR emitting height relation identified in a sample of disks around T~Tauri stars as part of the DARTTS-S program \citep{Avenhaus18} as a dashed red line in Figure \ref{fig:CO_13CO_vs_NIR_Surface}.

The NIR surfaces lie between the CO and $^{13}$CO emitting layers in HD~163296, while in IM~Lup, the NIR surface appears at approximately the same height as that of the $^{13}$CO, which is roughly consistent with the findings from \citet{Pinte18} and \citet{Rich21}. Although lacking well defined rings, MWC~480 has been reported to have a very flat NIR surface, i.e., z/r$\sim$0.03 \citep{Kusakabe12}, which suggests that the micron-sized dust lies at or below the $^{13}$CO and C$^{18}$O emitting layers.

\subsection{Gas temperatures}

We can use line emitting surfaces together with line brightness temperatures to map disk temperature structures. When we extracted individual pixels from the image cubes, we also obtained a corresponding set of peak surface brightnesses. In Subsection \ref{sec:calc_gas_temp}, we describe how we converted these peak surface brightnesses into gas temperatures as a function of ($r$, $z$). Then, in Subsection \ref{sec:radial_temp}, we present the radial temperature profiles and in Subsection \ref{sec:2d_temperature_str}, we analyze the full 2D empirical temperature structure of each disk. Both the radial temperature profiles and full ($r$, $z$) temperature structures for each MAPS are provided as publicly-available VADPs (see Section \ref{sec:VADPs}).

\subsubsection{Calculating Gas Temperatures} \label{sec:calc_gas_temp}

The peak of the CO and $^{13}$CO 2--1 lines are expected to be optically thick with CO rotational levels in local thermodynamic equilibrium \citep[e.g,][]{Weaver18} at the typical densities and temperatures of protoplanetary disks. Provided that the emission fills the beam, the peak surface brightness I$_{\nu}$ provides a measure of the temperature of the emitting gas. In order to not underestimate the line intensity along lines of sight containing optically thick continuum emission \citep[e.g,][]{Boehler17, Weaver18}, we repeated the surface fitting, as in Section \ref{sec:deriving_emission_surfaces}, using the non-continuum-subtracted image cubes \citep{czekala20}.

Each individual pixel ($r$, $z$) that was extracted has a peak surface brightness $I_{\nu}$, which was then used to calculate the associated gas temperature using the full Planck function:

\begin{equation} \label{eqn:tb}
T_b = \frac{h \nu}{k_B} \left[ \ln \left( \frac{2 h \nu^3 }{c^2 I_{\nu}} + 1 \right) \right]^{-1}
\end{equation}

In addition to CO 2--1 and $^{13}$CO 2--1, we also calculated the brightness temperatures of C$^{18}$O 2--1 in all disks and those of HCN 3--2 and C$_2$H 3--2 in HD~163296, but as we expect these lines to be partly optically thin, their brightness temperatures will be lower limits on the gas temperatures.

The western half of the AS~209 disk suffers from foreground cloud contamination \citep{Oberg11_DISCS} in CO 2--1. Therefore, we calculated CO 2--1 temperatures using only the eastern half of the disk, which corresponds to the velocity range of 4.90~km~s$^{-1}$ to 6.90~km~s$^{-1}$ (see Appendix \ref{sec:app:isovelocity}), to avoid underestimating the peak brightness temperatures. For all other lines, we used the same velocity/channel ranges as in Table \ref{tab:emission_surf} for temperature calculations. 

All subsequent radial and 2D gas temperature distributions represent those derived directly from individual surface measurements, rather than radially-deprojecting peak intensity maps \citep[see][]{teague20} or mapping peak brightness temperatures back onto derived emission surfaces (i.e., Figure \ref{fig:Surface_Moment_Overlay}). We only consider the brightness temperatures of those pixels where we were able to determine an emission height.

\subsubsection{Radial temperature distributions} \label{sec:radial_temp}

\begin{figure*}[ht]
\centering
\includegraphics[width=\linewidth]{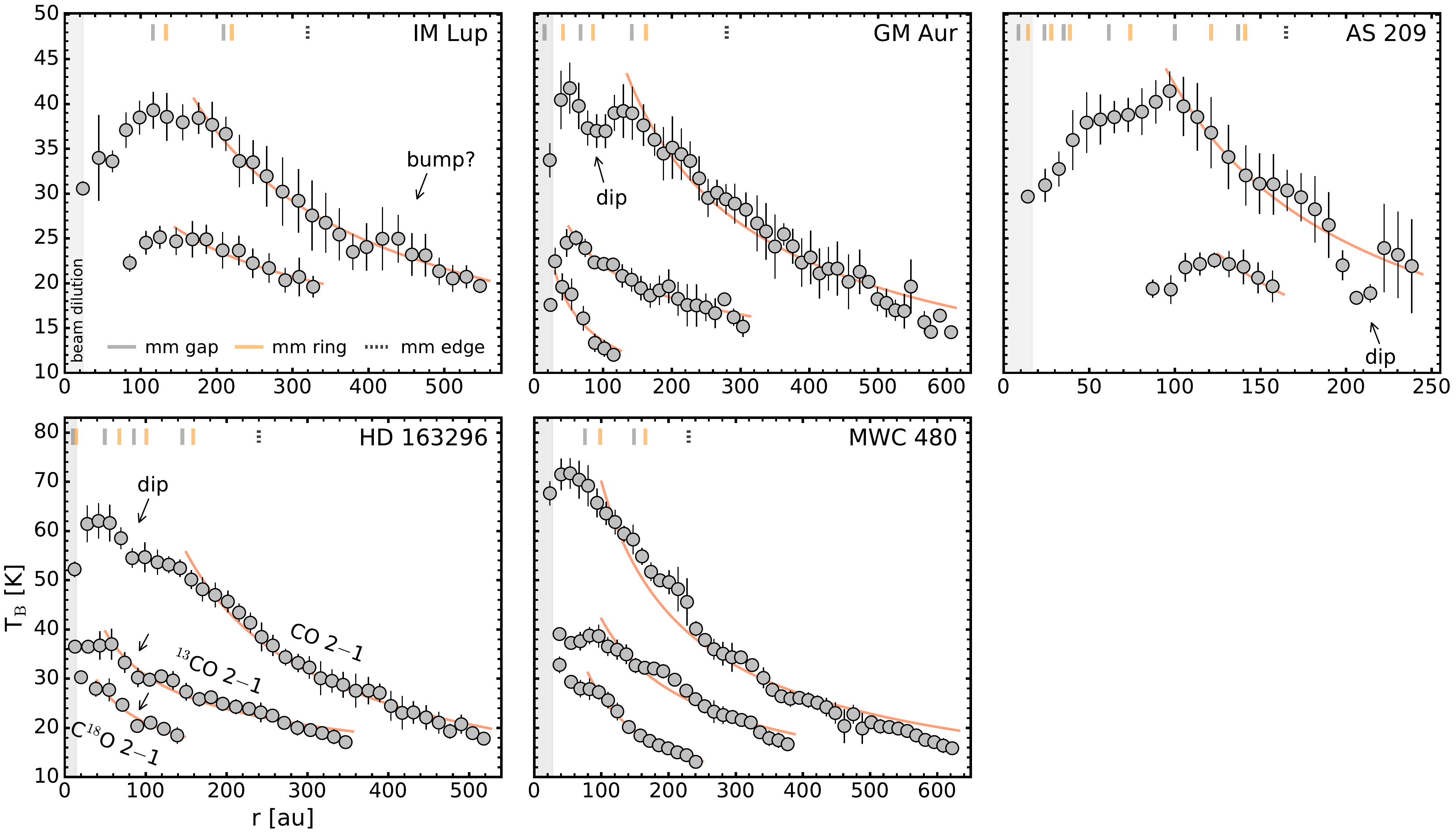}
\caption{Radial brightness temperature profiles for CO 2--1, $^{13}$CO 2--1, and when available, C$^{18}$O 2--1. The top and bottom rows are each shown on a consistent temperature scale and are grouped by whether the host star is a T~Tauri or Herbig~Ae star, respectively. These profiles represent the mean temperatures computed by radially binning the individual measurements, similar to the procedure used to compute the radially-binned surfaces (see Section \ref{sec:calc_gas_temp}). Vertical lines show the 1$\sigma$ uncertainty, given as the standard deviation of the individual measurements in each bin. For increased visual clarity, all disks have been binned by an extra factor of two, except for IM~Lup and HD~163296, which have been binned by an additional factor of three and four, respectively. The solid red line shows the fitted power law profile from Table \ref{tab:radial_temperature_plaw_fits}. The inner gray shaded region is the FWHM of beam major axis. The orange and gray dashes correspond to the mm continuum rings and gaps, respectively. The black dots mark the outer edge of the millimeter continuum. Temperature dips in GM~Aur, AS~209, and HD~163296 are labeled with arrows, as is a temperature bump in IM~Lup.}
\label{fig:Tb_CO_B6_profiles}
\end{figure*}

Figure \ref{fig:Tb_CO_B6_profiles} shows the radial temperature profiles for the CO isotopologues. We first reiterate that these temperatures are measurements of surface brightnesses, rather than integrated intensities that are used to identify line emission substructures \citep[][]{law20b} or derive column densities \citep[][]{zhang20}. As expected, in each disk, CO is the warmest, followed by $^{13}$CO and then C$^{18}$O. CO displays the largest range of measured temperatures, while $^{13}$CO and C$^{18}$O span a more limited range. The radial temperature gradients are consistent within each disk with similar slopes across CO isotopologues with the exception of AS~209, where the $^{13}$CO is nearly flat over the entire radial range in which it was measured. This flatness in temperature structure is due to the $^{13}$CO emission rings in AS~209 and in this case, we are only able to derive the brightness temperature of the outer ring at ${\sim}$120~au \citep[][]{Favre19,law20b}.

The disks around T Tauri stars have brightness temperatures spanning ${\sim}$10--40~K. The disks around Herbig Ae stars HD~163296 and MWC~480 are generally warmer at a given radius and have an overall larger total temperature range from ${\sim}$10--70~K. The CO 2--1 temperatures are about 10~K higher in MWC~480 than in HD~163296, with the greatest differences occurring within ${<}$200~au. In particular, the HD~163296 and MWC~480 profiles are consistent with those presented in \citet{teague20}, which were instead generated by deprojecting the peak intensity maps rather than direct extraction from emitting surfaces. Likewise, the CO 2--1 temperature profile of HD~163296 is approximately consistent with, although slightly cooler than, the one derived from a similar direct extraction method in \citet{Isella18}. We note that brightness temperatures less than 20~K are below the CO freeze-out temperature, which suggests that the associated line emission is at least partially optically thin and thus only provides a lower limit on the true gas temperatures. This conclusion is supported by our data, where CO lines with ${<}$20~K are most common for the rarer isotopologues and at large disk radii.

The drop in brightness temperature seen within 20--40~au in all disks and lines, which is marked as a shaded region in Figure \ref{fig:Tb_CO_B6_profiles}, is due to beam dilution as the emitting area becomes comparable to or smaller than the angular resolution of the observations. In the case of IM~Lup and AS~209, the central temperature dip extends further than the beam size. This may suggest enough CO depletion for the lines to become optically thin at these innermost radii, unresolved CO emission substructure, or that a substantial fraction of the CO emission is absorbed by dust. Indeed, \citet{Cleeves16} and \citet{sierra20} find that the dust is optically thick in the inner regions of the IM~Lup disk, and \citet{bosman20_inner_au} also see a large CO emission gap which is best explained by dust absorption. Dust absorption may also contribute to the low CO temperature in the inner AS~209 disk, but not out to 100~au. Optically thin emission is also an unlikely explanation: \citet{zhang20} finds that while AS~209 has a lower CO surface density than all other MAPS disks, it is still far from the optically thin limit in CO 2-1. This leaves CO substructure as an explanation. AS~209 does present several gaps in CO emission interior to 100~au \citep{Guzman18,law20b,zhang20,bosman20_inner_au}, which are barely resolved and may therefore result in a low brightness temperature in the inner disk.

Figure \ref{fig:Tb_CO_B6_profiles} also shows the locations of millimeter continuum gaps and rings, as reported in \citet{law20b}. The radial temperature profiles are quite smooth and hence the opportunity for coincidences between temperature substructures and other substructures is small. In two cases, the temperature substructure that is seen does line up with known disk substructures, however. HD~163296 has a 5~K drop in temperature at ${\sim}$80-90~au in all three CO lines, which aligns with a gap at 85~au in the millimeter continuum. A similar drop in temperature at ${\sim}90$~au is also present in CO in GM~Aur, which roughly aligns with a gap-ring pair at 68~au and 86~au in the millimeter continuum. In AS~209, a slightly deeper (${\sim}$8~K) drop occurs at 200~au in CO 2--1 and is coincident with a CO 2--1 line emission gap at 197~au \citep{law20b}. Low-amplitude (${\sim}$2-3~K) wave-like fluctuations are seen in CO 2--1 temperature in MWC~480 \citep[for further discussion of the features, see][]{teague20}. We find no association between temperature trends and the outer continuum edge in GM~Aur and AS~209, but do notice a modest flattening of the CO temperature gradient at the edge of the millimeter continuum in HD~163296 and MWC~480. Although about 50~au beyond the continuum edge, the CO 2--1 temperature in IM~Lup shows a modest increase at 450~au. This may be associated with a temperature inversion in the midplane \citep{2016_inversion} and is broadly consistent with the radial location of 400~au predicted in the models of \citet{Facchini17}.

We fitted the temperature profiles with power laws as:

\begin{equation}
T = T_{100} \times \left(\frac{r}{\rm{100\,au}} \right)^{-q}
\end{equation}

We first visually chose the radial range in which the temperature profiles behave like a power law and then fitted each profile using the Levenberg-Marquardt minimization implementation in \texttt{scipy.optimize.curve\_fit}. The fitting ranges and derived parameters are listed in Table \ref{tab:radial_temperature_plaw_fits}.

\begin{deluxetable*}{lccccc}
\tablecaption{Radial Temperature Profile Fits\label{tab:radial_temperature_plaw_fits}}
\tablewidth{0pt}
\tablehead{
\colhead{Source} & \colhead{Line} & \colhead{r$_{\rm{fit, in}}$ [au]}  &\colhead{r$_{\rm{fit, out}}$ [au]} & \colhead{T$_{100}$ [K]} &  \colhead{q}} 
\startdata
IM Lup & CO 2$-$1 & 170 & 559 & 55~$\pm$~0.9 & 0.58~$\pm$~0.01\\
 & $^{13}$CO 2$-$1 & 145 & 339 & 30~$\pm$~0.6 & 0.32~$\pm$~0.03\\
GM Aur & CO 2$-$1 & 135 & 613 & 52~$\pm$~0.9 & 0.61~$\pm$~0.02\\
 & $^{13}$CO 2$-$1 & 50 & 314 & 22~$\pm$~0.2 & 0.26~$\pm$~0.01\\
 & C$^{18}$O 2$-$1 & 30 & 126 & 14~$\pm$~0.4 & 0.38~$\pm$~0.05\\
AS 209 & CO 2$-$1 & 95 & 244 & 42~$\pm$~1.0 & 0.78~$\pm$~0.05\\
 & $^{13}$CO 2$-$1 & 125 & 163 & 28~$\pm$~1.3 & 0.80~$\pm$~0.13\\
HD 163296 & CO 2$-$1 & 150 & 527 & 78~$\pm$~1.0 & 0.82~$\pm$~0.01\\
 & $^{13}$CO 2$-$1 & 50 & 356 & 31~$\pm$~0.2 & 0.37~$\pm$~0.01\\
 & C$^{18}$O 2$-$1 & 40 & 148 & 21~$\pm$~0.3 & 0.37~$\pm$~0.03\\
MWC 480 & CO 2$-$1 & 100 & 632 & 70~$\pm$~1.1 & 0.69~$\pm$~0.02\\
 & $^{13}$CO 2$-$1 & 100 & 388 & 42~$\pm$~0.9 & 0.60~$\pm$~0.03\\
 & C$^{18}$O 2$-$1 & 80 & 251 & 26~$\pm$~0.2 & 0.75~$\pm$~0.02\\
\enddata
\end{deluxetable*}

As shown in Figure \ref{fig:Tb_CO_B6_profiles}, these fits work well beyond ${\sim}$100--150~au in all disks but overpredict the measured brightness temperatures interior to this. The temperature profile of CO in MWC~480 changes slope at approximately 350~au, which complicates the choice of radial fitting range. It is possible to achieve a modestly more accurate fit if instead two power laws are used, one for the inner disk between 40 and 200~au and another for the outer disk from ${>}135$~au. However, for simplicity, we fit a single power to the maximal possible range.

\begin{figure}[ht]
\centering
\includegraphics[width=\linewidth]{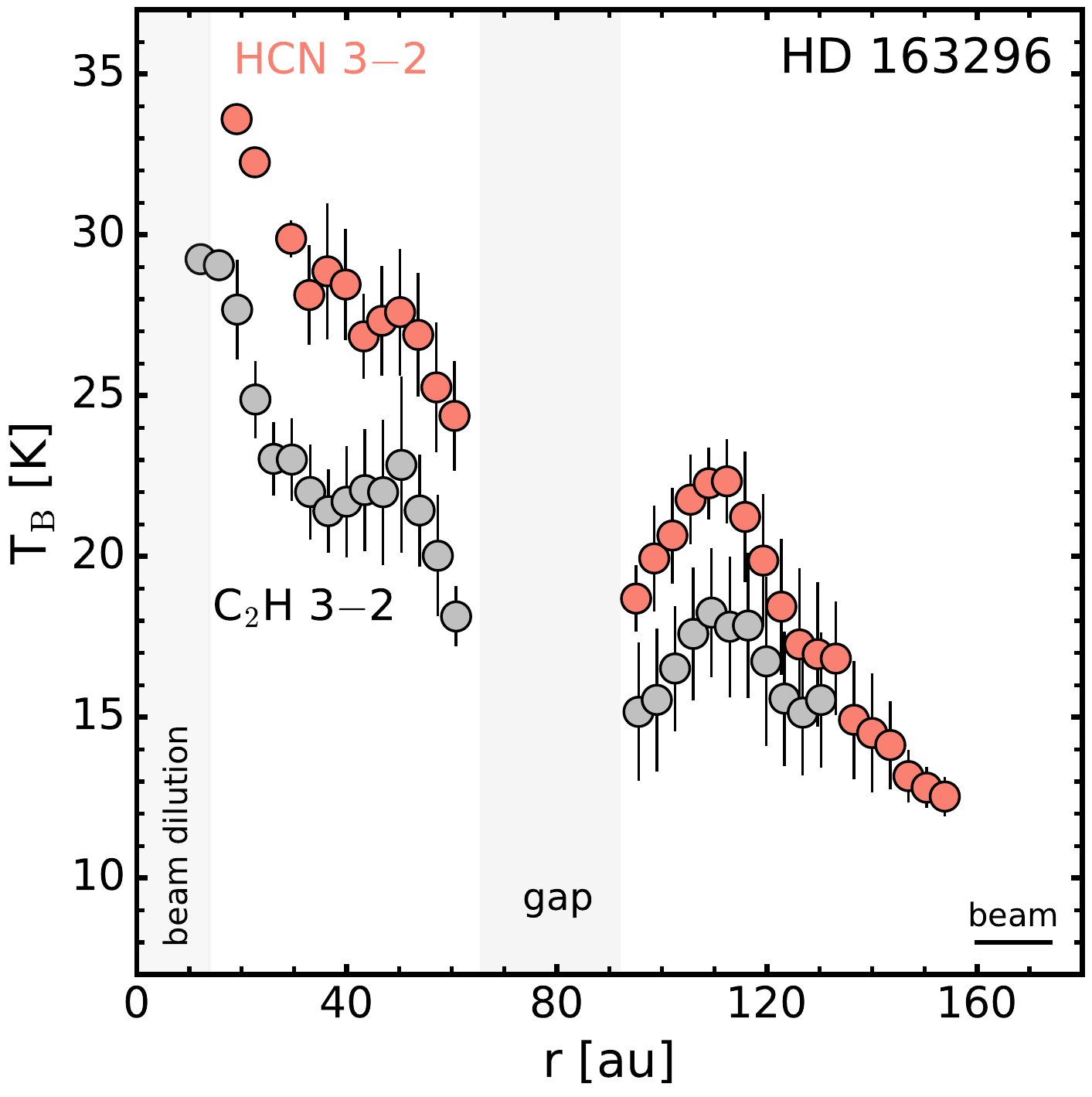}
\caption{Radial brightness temperature distributions of HCN 3--2 and C$_2$H 3--2 in HD~163296. These profiles represent the mean temperatures computed by radially binning the individual measurements, similar to the procedure used to compute the radially-binned surfaces (see Section \ref{sec:calc_gas_temp}). Vertical lines show the 1$\sigma$ uncertainty, given as the standard deviation of the individual measurements in each bin. The gap location seen in the line emission \citep{law20b} is labeled and shaded in gray. The FWHM of the beam major axis is shown in the lower right corner and also indicated by the inner gray shaded region.}
\label{fig:Tb_C2H_HCN_HD16}
\end{figure}

In addition to the CO lines, we also derived the brightness temperature profiles of HCN and C$_2$H in HD~163296, as shown in Figure \ref{fig:Tb_C2H_HCN_HD16}. The shapes of the temperature profiles are consistent but offset, as HCN is warmer by 4--6~K at all radii. The gap at ${\sim}$80~au has a C$_2$H temperature of ${<}15$~K and HCN temperature of ${<}20$~K. Both lines seem to be cooler near the gap by a few K relative to the outer ring and by almost 10~K versus the inner ring. However, the beam filling factor will be reduced at locations closer to the gap, likely becoming significant within 1/2 -- 1 beams away. Thus, the line emission near this gap may become increasingly optically thin. In this case, the lower brightness temperatures would reflect reduced gas density, rather than cooler HCN and C$_2$H gas temperatures. Overall, the HCN brightness temperatures are consistent with the excitation temperatures derived in the multi-line analysis of \citet{guzman20}. However, the C$_2$H temperatures are a factor of two lower than those reported in \citet{guzman20}, which suggests that the C$_2$H 3--2 line is optically thin or not in local thermal equilibrium (LTE). Thus, a non-LTE analysis of C$_2$H in HD~163296 is warranted.

\subsubsection{2D temperature structure} \label{sec:2d_temperature_str}

\begin{figure*}[!ht]
\centering
\includegraphics[width=\linewidth]{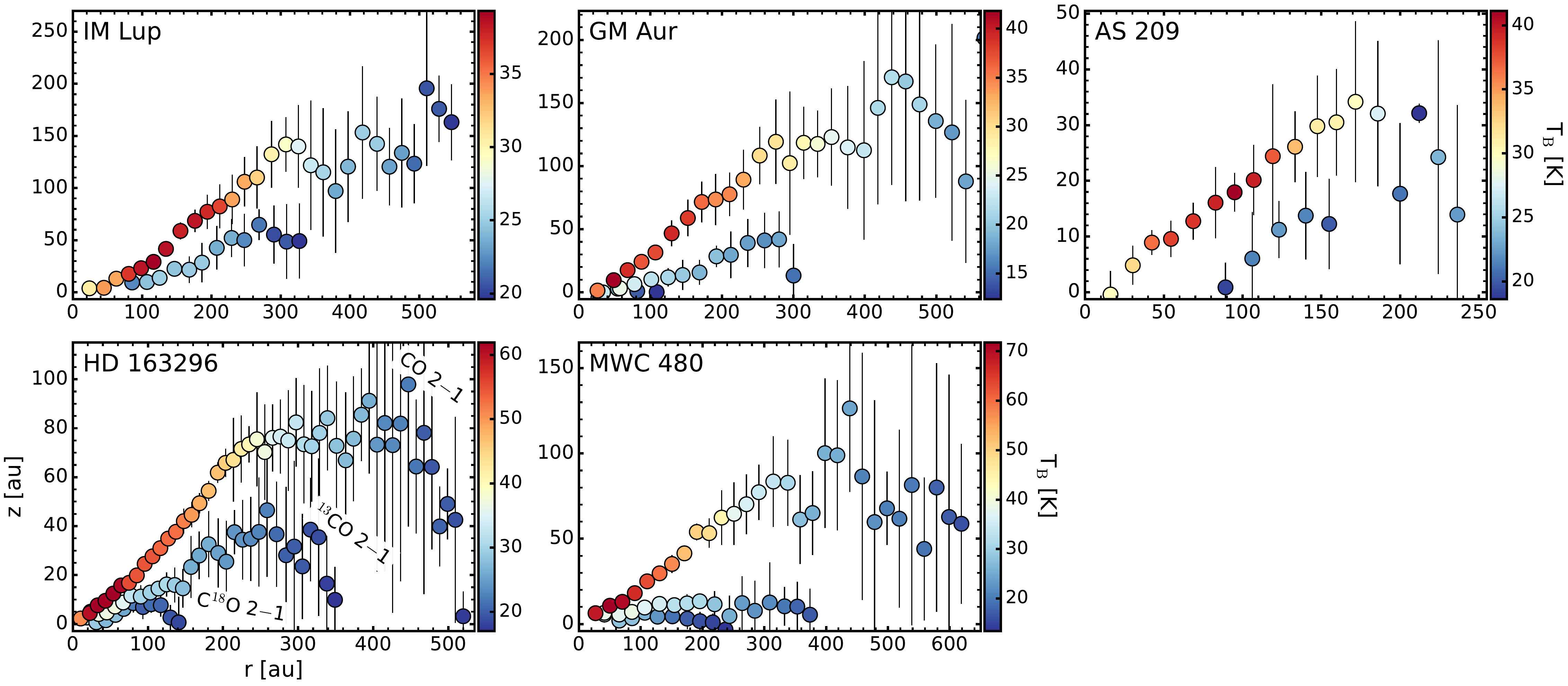}
\caption{2D temperature distributions of CO 2--1, $^{13}$CO 2--1, and when available, C$^{18}$O 2--1 in all MAPS sources. Points are those from the binned surfaces and error bars are the 1$\sigma$ uncertainties in $z$. For some of the innermost points, the uncertainty is smaller than the marker. Data for all disks and lines have been binned by a factor of three for visual clarity. The uncertainty of the temperature measurements, which is not shown here, can be found in Figure \ref{fig:Tb_CO_B6_profiles}.}
\label{fig:2D_Temperature_Data}
\end{figure*}

The advantage of having multiple CO isotopologues that trace different disk heights is access to the vertical temperature distribution. \citet{Dartois_ea_2003} were the first to demonstrate this in moderate resolution (${\sim}$1$^{\prime \prime}$) observations of DM~Tau. More recently, \citet{Pinte18} presented a framework for directly mapping the temperature structure of each emitting layer in (${\sim}$0\farcs4) observations of IM~Lup. Here, we expanded this analysis to the high spatial resolution observations of the MAPS disks. Figure \ref{fig:2D_Temperature_Data} shows the full 2D temperature distributions.

Since we have temperature information as a function of $(r\,,z)$, we can construct a full 2D model of the temperature distribution of each disk. To do so, we adopt a two-layer model similar to the one proposed by \citet{Dartois_ea_2003}, but then modified by \citet{Dullemond20} with a different connecting term. Both formula were initially tried but substantially better fits were obtained with that of \citet{Dullemond20}. The midplane temperature $T_{\rm mid}$ and atmosphere temperature $T_{\rm atm}$ are assumed to have a power-law profile with slopes $q_{\rm mid}$ and $q_{\rm atm}$, respectively.

\begin{equation}
T_{\rm{atm}} (r) = T_{\rm{atm}, 0} \left( r / 100~\rm{au} \right)^{q_{\rm{atm}}}
\end{equation}

\begin{equation}
T_{\rm{mid}} (r) = T_{\rm{mid}, 0} \left( r / 100~\rm{au} \right)^{q_{\rm{mid}}}
\end{equation}

Between the midplane and atmosphere, the temperature is smoothly connected using a tangent hyperbolic function

\begin{equation} \label{eqn:trig}
T^4 (r, z) =  T^4_{\rm{mid}} (r) + \frac{1}{2} \left[ 1 + \tanh \left( \frac{z - \alpha z_q(r)}{z_q(r)} \right) \right] T^4_{\rm{atm}} (r),
\end{equation}

\noindent where $z_q (r) = z_0 \left(r / 100~\rm{au} \right)^{\beta}$. We note that the $\alpha$ parameter defines where in height the transition in the tanh vertical temperature profile occurs and $\beta$ describes how the transition height varies over radius. In total, we fitted the following seven parameters: $T_{\rm atm,0}$, $q_{\rm atm}$, $T_{\rm mid,0}$, $q_{\rm mid}$, $\alpha$, $z_0$, and $\beta$.

We performed the fitting using MCMC with \texttt{emcee} \citep{Foreman_Mackey13} with 256 walkers which take 500 steps to burn in and an additional 5000 steps to sample the posterior distribution function. All available CO lines were fitted using the individual measurements and only those points with $T_{\rm{B}}>20$~K were considered. Temperatures below 20~K, close to the CO freeze-out temperature, are likely optically thin and thus not useful for constraining the gas temperature structure. Parameter values and associated uncertainties are taken to be the 50th, 16th, and 84th percentiles from the marginalized posterior distributions, respectively, and are listed in Table \ref{tab:2D_temperature_fit_params}.

\begin{deluxetable*}{lccccccccccccc}
\tablecaption{Summary of 2D Temperature Structure Fits\label{tab:2D_temperature_fit_params}}
\tablewidth{0pt}
\tablehead{\colhead{Source} & \colhead{$T_{\rm{atm},0}$ [K]} & \colhead{$T_{\rm{mid}, 0}$ [K]} & \colhead{$q_{\rm{atm}}$}& \colhead{$q_{\rm{mid}}$} & \colhead{$z_0$ [au]} & \colhead{$\alpha$} & \colhead{$\beta$} }
\startdata
IM Lup & 36$^{+0.1}_{-0.1}$ & 25$^{+0.1}_{-0.1}$ & 0.03$^{+0.01}_{-0.01}$ & $-$0.02$^{+0.01}_{-0.01}$ & 3$^{+0.1}_{-0.1}$ & 4.91$^{+0.17}_{-0.16}$ & 2.07$^{+0.02}_{-0.02}$\\
GM Aur & 48$^{+0.3}_{-0.3}$ & 20$^{+0.2}_{-0.2}$ & $-$0.55$^{+0.01}_{-0.01}$ & $-$0.01$^{+0.01}_{-0.01}$ & 13$^{+0.2}_{-0.2}$ & 2.57$^{+0.03}_{-0.03}$ & 0.54$^{+0.01}_{-0.01}$\\
AS 209 & 37$^{+0.2}_{-0.2}$ & 25$^{+0.2}_{-0.2}$ & $-$0.59$^{+0.02}_{-0.02}$ & $-$0.18$^{+0.01}_{-0.01}$ & 5$^{+0.2}_{-0.2}$ & 3.31$^{+0.12}_{-0.11}$ & 0.02$^{+0.02}_{-0.02}$\\
HD 163296 & 63$^{+0.2}_{-0.2}$ & 24$^{+0.1}_{-0.1}$ & $-$0.61$^{+0.003}_{-0.003}$ & $-$0.18$^{+0.004}_{-0.004}$ & 9$^{+0.1}_{-0.1}$ & 3.01$^{+0.02}_{-0.02}$ & 0.42$^{+0.004}_{-0.004}$\\
MWC 480 & 69$^{+0.2}_{-0.2}$ & 27$^{+0.2}_{-0.2}$ & $-$0.7$^{+0.004}_{-0.004}$ & $-$0.23$^{+0.01}_{-0.01}$ & 7$^{+0.1}_{-0.1}$ & 2.78$^{+0.02}_{-0.02}$ & $-$0.05$^{+0.01}_{-0.01}$\\
\enddata
\end{deluxetable*}

\begin{figure*}[!ht]
\centering
\includegraphics[width=\linewidth]{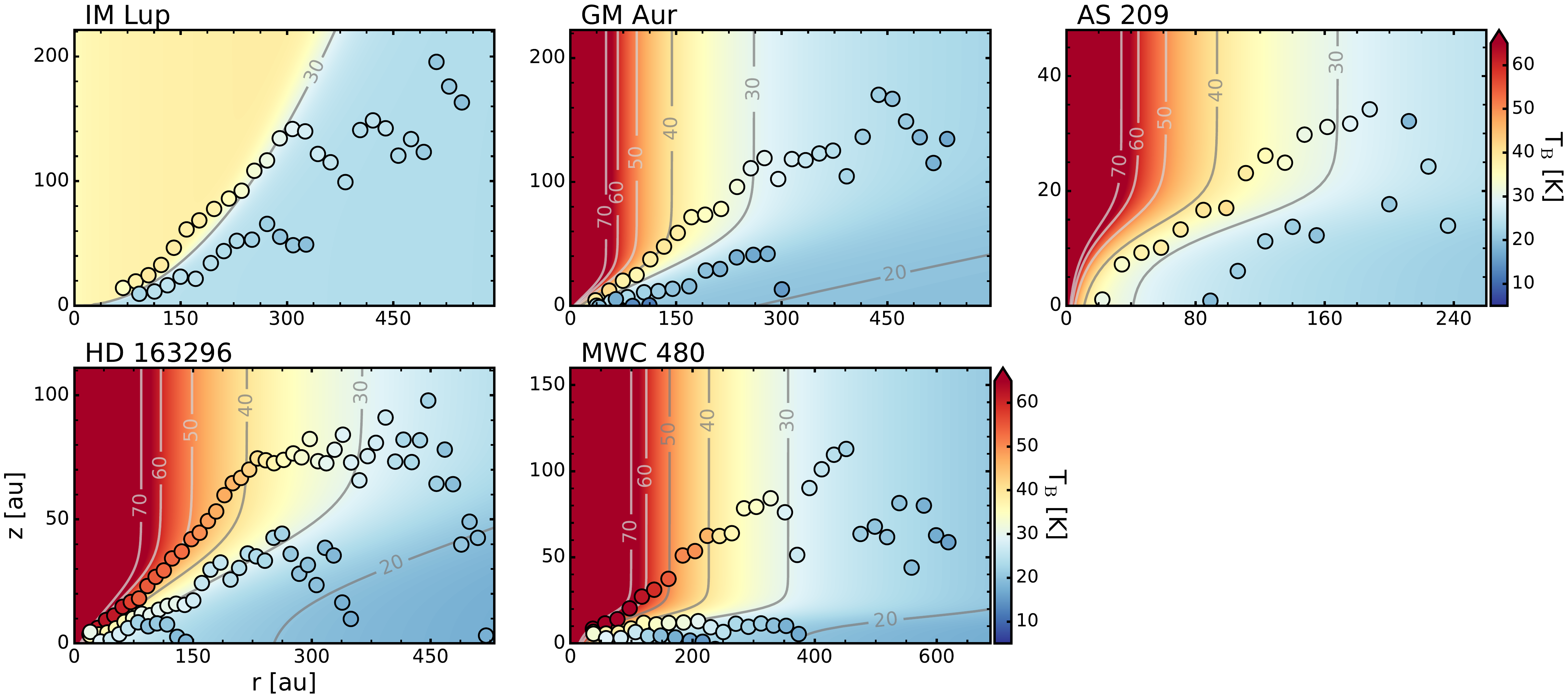}
\caption{Comparison of the measured temperatures (points) with the fitted 2D temperature structures (background), as listed in Table \ref{tab:2D_temperature_fit_params}. The same color scale is used for the data and fitted model and is consistent across all panels. Contours show constant temperatures in increments of 10~K. Data for all disks and lines have been binned by a factor of three for visual clarity. The uncertainty of the temperature measurements, which is not shown here, can be found in Figure \ref{fig:Tb_CO_B6_profiles}.}
\label{fig:2D_Temperature_Fits}
\end{figure*}

The 2D fitted models are shown in comparison with the data in Figure \ref{fig:2D_Temperature_Fits}. For all disks, the median residuals between the fitted model and measured temperatures are typically no more than 10\%. The most informative fits are those with a well-sampled $(r,\,z)$ space, which means that we have a set of CO isotopologue lines with a diverse set of $z/r$ values, e.g., HD~163296. In contrast, IM~Lup is poorly constrained over the height of the disk, since surfaces were only able to be determined for CO and $^{13}$CO and they are not widely spaced in $z/r$. The abrupt change in $z/r$ from CO to $^{13}$CO and C$^{18}$O in MWC~480 is also reflected in its inferred 2D temperature structure by its small fitted $\alpha$ and $\beta$ values (Table \ref{tab:2D_temperature_fit_params}). This means that the transition in vertical temperature, as described in Equation \ref{eqn:trig}, occurs close to the midplane and the transition height does not increase over radius, unlike other disks. In general, as the emitting surfaces do not provide direct constraints in the disk midplanes, we caution the use of the empirically derived $T_{\rm{mid}}$, which are considerably warmer than predictions from thermo-chemical models \citep{zhang20}.

\subsection{Substructures in emission surfaces} \label{sec:substructures_emisssion_surfaces}

Localized vertical substructures are observed in many of the emission surfaces derived from the MAPS data. The properties of these substructures, namely their radial locations, widths, and depths, provide important constraints that are necessary for detailed thermo-chemical modelling \citep[e.g.,][]{Rab20,calahan20b}. In the following subsections, we identify and catalogue all substructures present in the derived emitting layers and compare them with the gas temperature profiles, and with substructures observed in the millimeter continuum and CO line emission.

\subsubsection{Fitting vertical substructures} \label{sec:fitting_vertical_substr}
Each substructure is labeled with its radial location rounded to the nearest whole number in astronomical units and is preceded with ``Z'' to indicate these features are vertical variations. This nomenclature is also chosen to avoid ambiguity with that used to denote radial substructures in the continuum \citep{Huang18} and molecular line emission  \citep{law20b} profiles, which labels rings by ``B" (``bright'') and gaps by ``D" (``dark").

\begin{figure}[!ht]
\centering
\includegraphics[width=\linewidth]{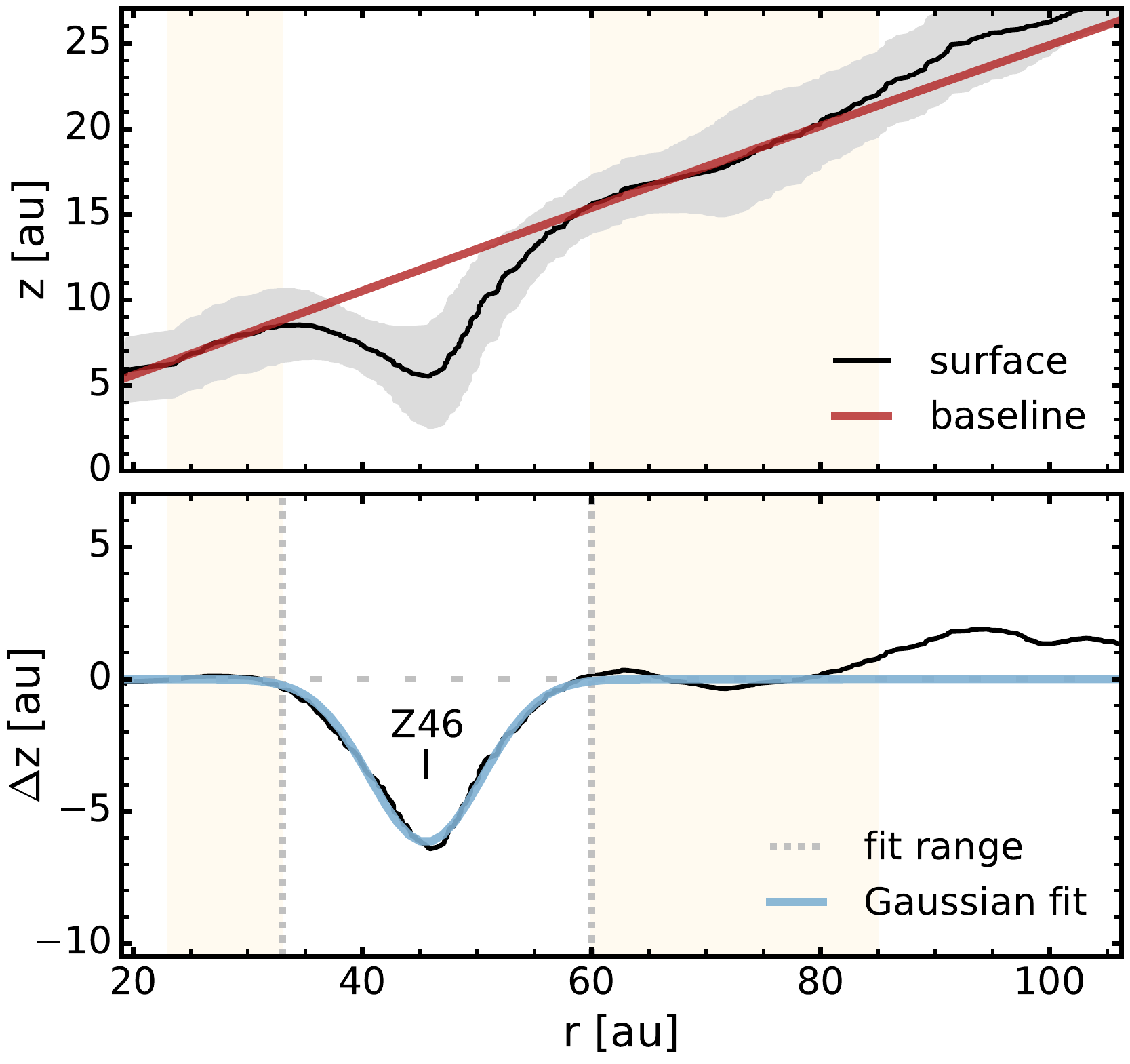}
\caption{Example of the vertical substructure fitting process in the CO 2--1 emission surface of HD~163296. The top panel shows the removal of a local baseline (solid red line) in the form of a quadratic polynomial fit. The bottom panel shows a Gaussian fit (solid blue line) to the Z46 substructure in the baseline-subtracted emission surface. The radial range for the local baseline was visually identified and is shown by the orange shaded regions.}
\label{fig:fitting_vertical_substructures}
\end{figure}

Feature identification was done visually and focused on the inner, rising portion of the surfaces within ${\sim}$200~au, which was the most well-constrained and possessed the highest SNRs. We used the moving average surfaces to search for substructures in the form of vertical dips, i.e., we assumed that substructures represent localized decreases in $z$ in an otherwise smoothly-varying emitting layer. To fit each substructure, we first visually estimated a local baseline. This baseline was then fitted with a quadratic polynomial and subtracted from the original emitting surface. The derived properties of each feature will depend on the assumed form of the local baseline, but a low-order polynomial baseline is sensible for the inner ${<}200$~au of each disk. We then fitted a Gaussian profile to characterize each feature in the baseline-subtracted surface. An example of this fitting process for CO 2--1 in HD~163296 is shown in Figure \ref{fig:fitting_vertical_substructures}.

The fitted centers and FWHMs of each Gaussian are taken to be the radial location and widths of each feature, respectively. Substructure depths are defined as $\Delta z$/$z_{\rm{baseline}}$, where $z_{\rm{baseline}}$ is the vertical height of the fitted baseline and $\Delta z$ is the fitted vertical height of the emitting surface at the radial position of the substructure. Depths are subsequently referred to according to their fractional decrease in vertical height with deeper features having lower height ratios, e.g., the Z46 in HD~163296 has a $\Delta z$/$z_{\rm{baseline}}=0.52$, which indicates a depth of 48\%. The center, width, and relative depth of each feature is listed in Table \ref{tab:vertical_substructures} and their radial locations are labeled in Figure \ref{fig:mm_NIR_comparison_1}.

The relative depth of each feature is sometimes more uncertain than Table \ref{tab:vertical_substructures} suggests, as depth strongly depends on the assumed baseline, but overall, we find that this method works well to identify substructure radial locations and provides a preliminary characterization. These definitions also do not explicitly account for beam effects. In a few cases, the widths and depths of individual features are smaller than the minor axis of the beam FWHM. However, this is not generally a concern, as surfaces are derived from the positional offsets of peak intensities, which are sensitive to scales smaller than the beam size.

Typical feature depths range from ${\sim}$30--70\% and widths from 10--50~au. HD~163296 has the largest total number of identified substructures, which are relatively narrow (${\sim}$10--15~au), while those in IM~Lup and MWC~480 have broader widths (${\sim}$30--50~au). A consistent broad, bowl-shaped depression is seen in all CO lines in MWC~480 around 66~au with a width of ${\gtrsim}$30--40~au and depth of 40--60\%. The two features Z170 and Z375 associated with IM~Lup are notable, as they occur at the largest radii of all identified substructures. Although having modest relative depths of 68\% and 49\%, they possess an absolute $\Delta$z of 46 and 18~au, which are the largest in au by a factor of a few to an order of magnitude compared to all other substructures. Features are not always present across all CO isotopologues. For instance, Z170 in IM~Lup is only seen in $^{13}$CO but not in CO, while in HD~163296, Z81 and Z83 are present in $^{13}$CO and C$^{18}$O, respectively, but no corresponding feature is identified in CO. 

In addition to the isolated, Gaussian-like dips we report above, we detect a few more complex trends. For instance, prominent changes in the slope of the emission surface are present at ${\sim}$150~au in $^{13}$CO in GM~Aur and at ${\sim}$115~au in CO in IM~Lup. Both AS~209 and MWC~480 also show large-scale, wave-like patterns in their CO surfaces with peak-to-trough separations of roughly 20~au and 40~au, respectively and amplitudes of no more than a few au. Thus, Z56 in AS~209 and Z66 in MWC~480 may in fact be local minima associated with this larger wave rather than separate isolated substructures.

\begin{deluxetable*}{lllccccccccccccc}
\tablecaption{Properties of Vertical Substructures \label{tab:vertical_substructures}}
\tablewidth{0pt}
\tablehead{\colhead{Source} & \colhead{Line} & \colhead{Feature} & \colhead{$r_0$ [arcsec]} & \colhead{$r_0$ [au]} & \colhead{Width [arcsec]} & \colhead{Width [au]} & \colhead{$\Delta$z [arcsec]} & \colhead{$\Delta$z [au]} & \colhead{Depth$^a$}}
\startdata
IM Lup & CO 2$-$1 & Z375 & 2.38 $\pm$ 0.03 & 375 $\pm$ 5 & 0.18 $\pm$ 0.09 & 29 $\pm$ 14 &  0.29 $\pm$ 0.1 &  46 $\pm$ 16 & 0.32 $\pm$ 0.28\\
 & $^{13}$CO 2$-$1 & Z170 & 1.08 $\pm$ 0.01 & 170 $\pm$ 2 & 0.18 $\pm$ 0.002 & 29 $\pm$ 0.4 &  0.11 $\pm$ 0.004 &  18 $\pm$ 1 & 0.51 $\pm$ 0.49\\
AS 209 & CO 2$-$1 & Z56 & 0.47 $\pm$ 0.02 & 56 $\pm$ 2 & 0.1 $\pm$ 0.01 & 13 $\pm$ 1 &  0.03 $\pm$ 0.02 &  4 $\pm$ 2 & 0.33 $\pm$ 0.18\\
HD 163296 & CO 2$-$1 & Z46 & 0.45 $\pm$ 0.01 & 46 $\pm$ 1 & 0.11 $\pm$ 0.01 & 12 $\pm$ 1 &  0.06 $\pm$ 0.01 &  6 $\pm$ 1 & 0.52 $\pm$ 0.13\\
 & $^{13}$CO 2$-$1 & Z49 & 0.48 $\pm$ 0.02 & 49 $\pm$ 2 & 0.1 $\pm$ 0.003 & 11 $\pm$ 0.3 &  0.02 $\pm$ 0.004 &  2 $\pm$ 0.4 & 0.36 $\pm$ 0.03\\
 &  & Z81 & 0.8 $\pm$ 0.001 & 81 $\pm$ 0.1 & 0.11 $\pm$ 0.03 & 12 $\pm$ 3 &  0.05 $\pm$ 0.03 &  5 $\pm$ 4 & 0.43 $\pm$ 0.24\\
 &  & Z145 & 1.43 $\pm$ 0.01 & 145 $\pm$ 1 & 0.1 $\pm$ 0.01 & 10 $\pm$ 1 &  0.02 $\pm$ 0.03 &  2 $\pm$ 3 & 0.33 $\pm$ 0.28\\
 & C$^{18}$O 2$-$1 & Z83 & 0.82 $\pm$ 0.004 & 83 $\pm$ 0.4 & 0.16 $\pm$ 0.02 & 16 $\pm$ 2 &  0.06 $\pm$ 0.001 &  6 $\pm$ 0.1 & 0.66 $\pm$ 0.02\\
MWC 480 & CO 2$-$1 & Z66 & 0.41 $\pm$ 0.01 & 66 $\pm$ 1 & 0.19 $\pm$ 0.04 & 31 $\pm$ 7 &  0.04 $\pm$ 0.003 &  7 $\pm$ 1 & 0.42 $\pm$ 0.06\\
 & $^{13}$CO 2$-$1 & Z66 & 0.41 $\pm$ 0.06 & 66 $\pm$ 9 & 0.28 $\pm$ 0.04 & 46 $\pm$ 7 &  0.03 $\pm$ 0.02 &  6 $\pm$ 4 & 0.63 $\pm$ 0.09\\
 & C$^{18}$O 2$-$1 & Z71 & 0.44 $\pm$ 0.03 & 71 $\pm$ 5 & 0.32 $\pm$ 0.06 & 51 $\pm$ 9 &  0.03 $\pm$ 0.003 &  4 $\pm$ 0.5 & 0.63 $\pm$ 0.55\\
\enddata
\tablenotetext{a}{Depth of vertical substructure, defined as the ratio of $\Delta z$ to z$_{\rm{baseline}}$ at $r_0$ (see Section \ref{sec:fitting_vertical_substr}.)}
\end{deluxetable*}

\subsubsection{Comparison with gas temperature} \label{sec:compare_with_gas_temperature}

Since we have estimates for the gas temperatures, we also searched for coincidences between vertical and temperature substructures, but only found a single one, i.e., Z81 in $^{13}$CO in HD~163296. In IM~Lup, the outer edge of the CO temperature plateau occurs around 180~au, which is roughly coincident with the Z170 feature seen in the $^{13}$CO surface. The CO 2--1 radial temperature profiles and emitting surfaces in MWC~480 also both show wave-like patterns, which are very roughly coincident in radial location. For further discussion of these features, see \citet{teague20}. Otherwise, no other temperature trends are identified in any of the CO lines in the MAPS disks. Thus, the empirical gas temperatures do not generally seem to be sensitive to the presence of surface substructures. This may be explained, in part, due to deeper layers in the disk being more isothermal than the upper layers. Thus, a small change in the emission height would look like a bigger dip in temperature for CO than $^{13}$CO, for instance. Moreover, since the scale height is proportional to $\sqrt{T}$, this means that significant temperature differences are required for noticeable changes in the scale height.

\subsubsection{Comparison with millimeter continuum and line emission substructures} \label{sec:mm_continuum_vs_chemical_substr}

The majority of the MAPS disks show at least some spatial links between vertical substructures and either continuum or radial substructure in CO line emission, as shown in Figure \ref{fig:mm_NIR_comparison_1}. Below, we consider each of these possible spatial associations.

\begin{figure*}[!ht]
\centering
\includegraphics[width=\linewidth]{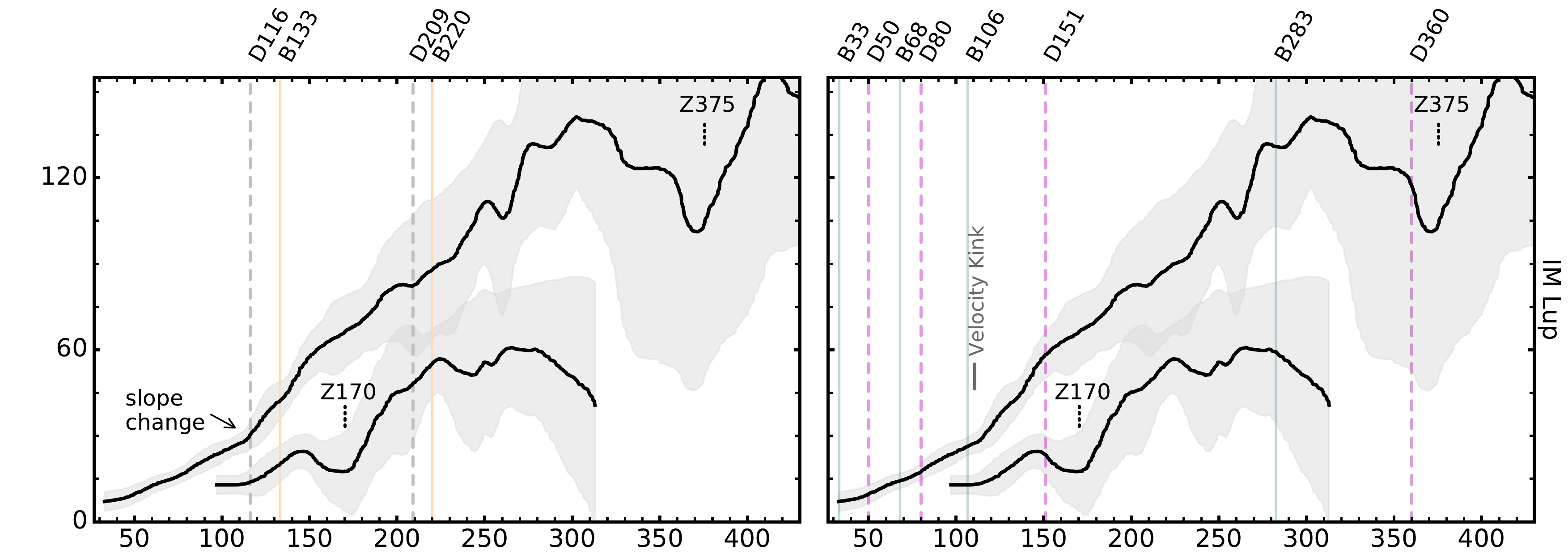}
\includegraphics[width=\linewidth]{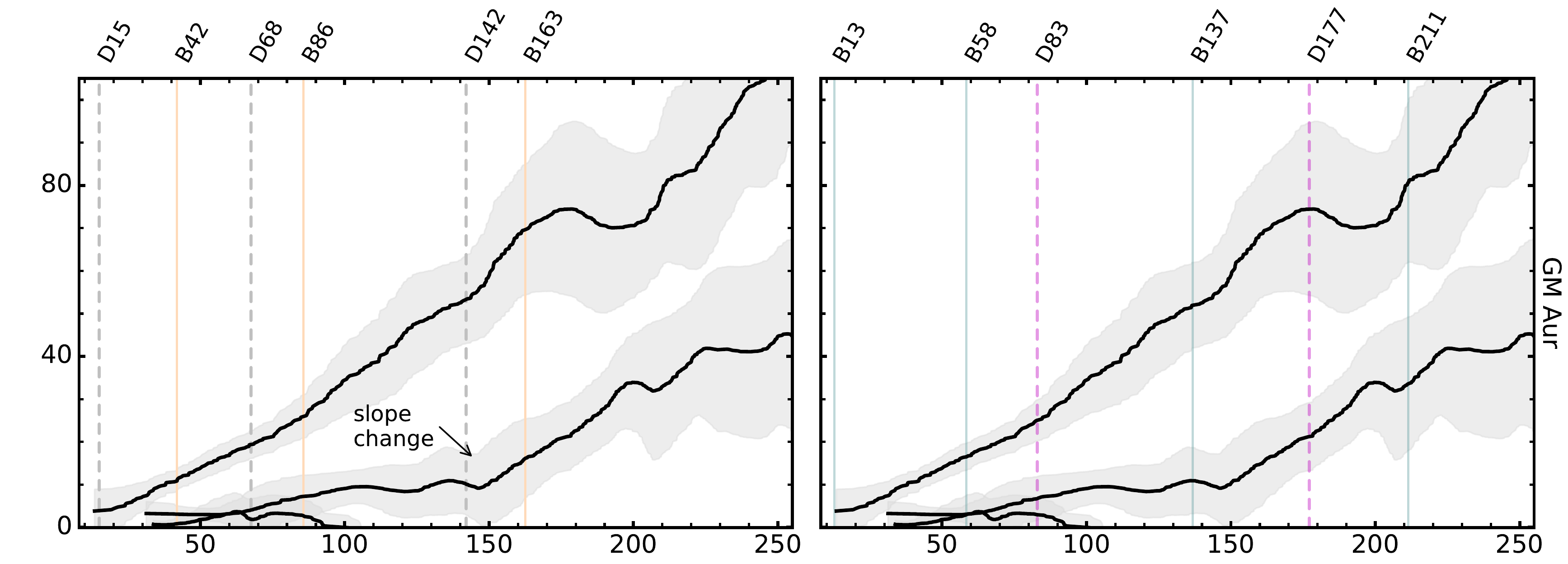}
\includegraphics[width=\linewidth]{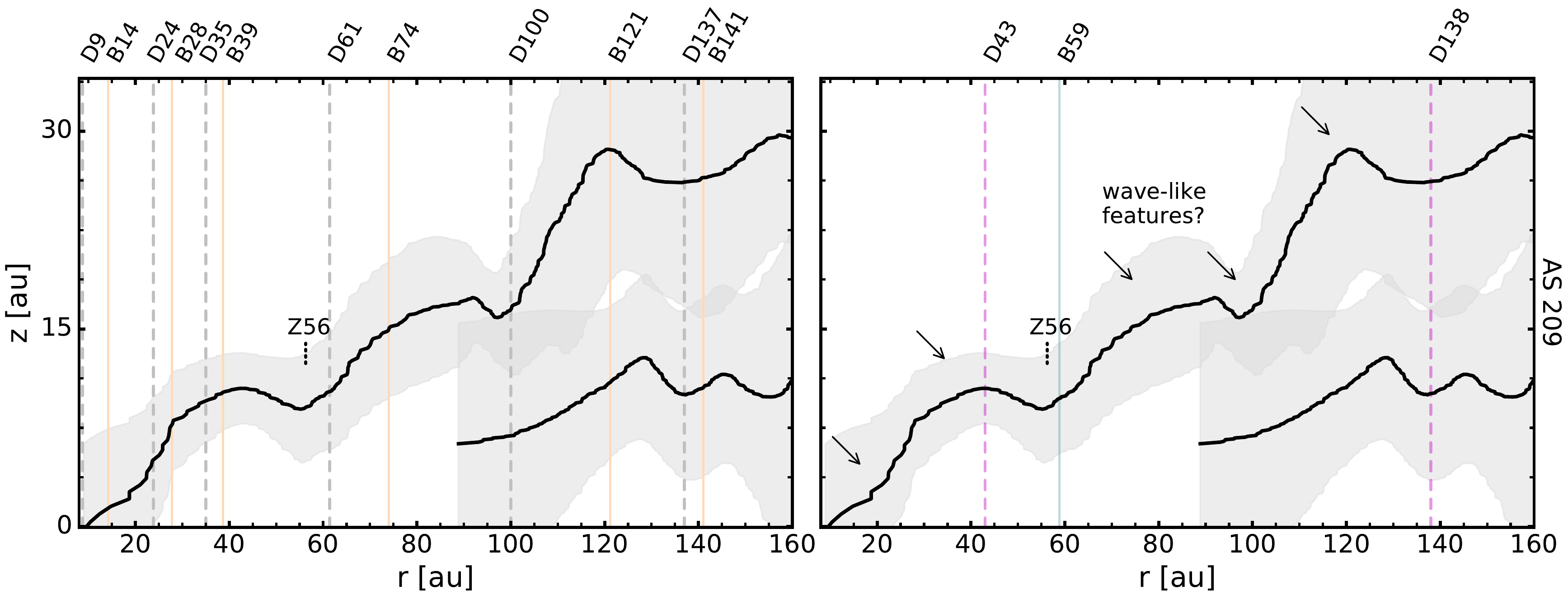}
\caption{Comparison of CO 2--1, $^{13}$CO 2--1, and C$^{18}$O 2--1 (when available) surfaces with annular substructures in the millimeter continuum (left) and CO line emission (right). Substructures are labeled following the nomenclature of \citet{Huang18} and taken from \citet{law20b}. The CO line emission substructures are from CO 2--1 for all disks, except for MWC~480, which show those from $^{13}$CO 2--1. Substructures are labeled as follows: dust ring (solid orange line), dust gap (dashed gray line), chemical ring (solid blue line), and chemical gap (dashed purple line). Dotted black lines mark vertical substructures, as listed in Table \ref{tab:vertical_substructures}. Notable changes in emitting surface slope and the suggestive wave-like features in AS~209 and MWC~480 are marked with arrows. The CO velocity kinks from \citet{Pinte18, Pinte20} are labeled in IM~Lup and HD~163296.}
\label{fig:mm_NIR_comparison_1}
\end{figure*}

\begin{figure*}[!ht]
\ContinuedFloat
\centering
\includegraphics[width=\linewidth]{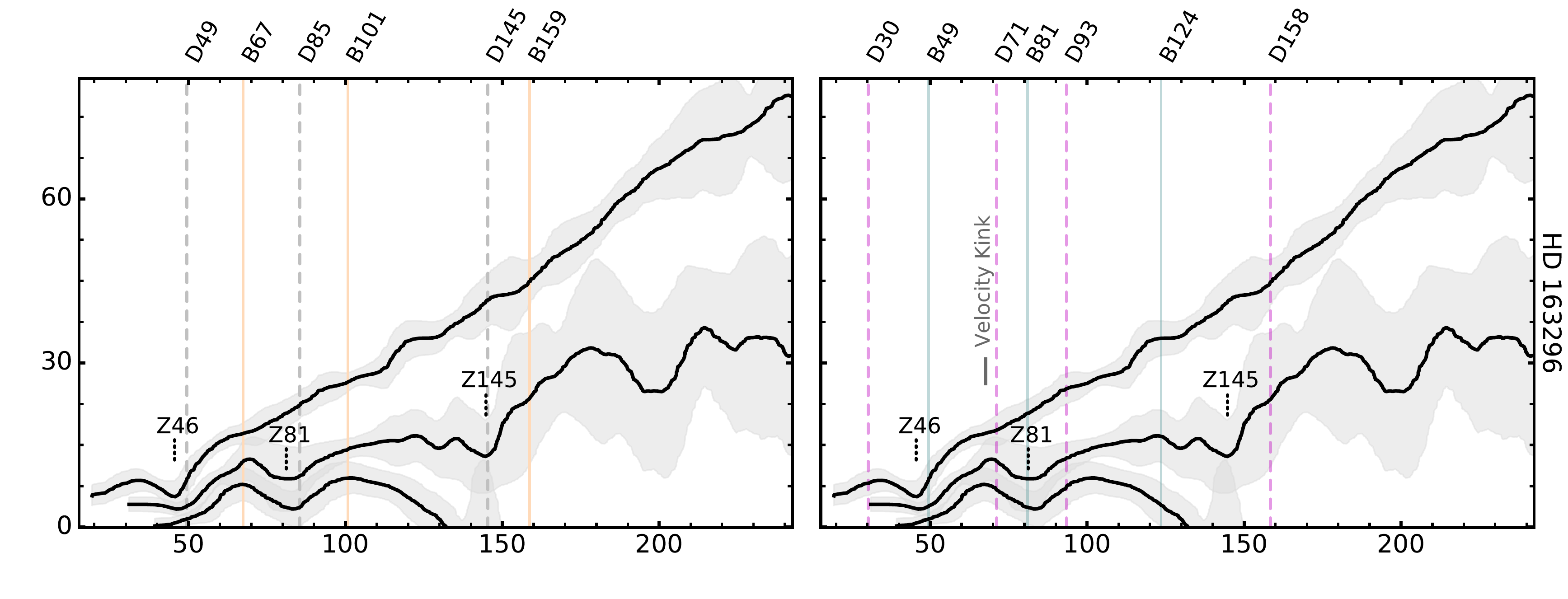}
\includegraphics[width=\linewidth]{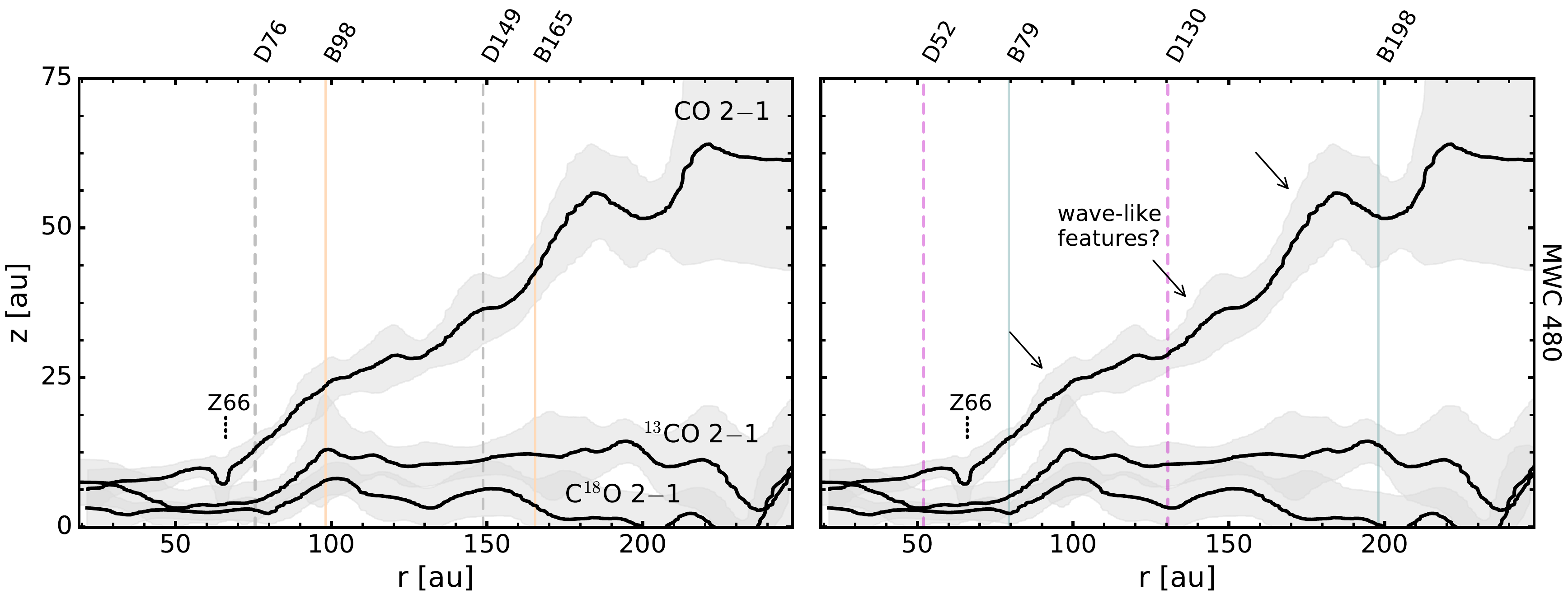}
\caption{ Continued.}
\label{fig:mm_NIR_comparison_2}
\end{figure*}

All MAPS disks show some degree of spatial association between surface features and millimeter continuum substructures. Each dust gap in HD~163296 aligns with a surface feature in at least one of the CO isotopologue surfaces and in the case of $^{13}$CO, there is a one-to-one match between millimeter gaps and surface substructures. In MWC~480, the inner dust gap D76 roughly aligns with the Z66 surface feature and the outer dust gap at D149 approximately matches the location of the 140~au trough of the wave-like fluctuations. However, these associations in MWC~480 are considerably more tentative and considering that the surface features are more than twice the width of continuum gaps, these may be chance alignments. Several of the wave-like surface features in AS~209 align with the radial locations of substructures in the millimeter continuum. In particular, Z56 and the troughs at ${\sim}$100 and 140~au are radially coincident with dust gaps. The changes in emitting surface slope in CO 2--1 in IM~Lup and $^{13}$CO 2--1 in GM~Aur are also both co-located with the D116 and D142 dust gaps, respectively.

In a few cases, we identify features in the CO line emission profiles that are radially coincident with vertical substructures. In HD~163296, the CO line emission peaks at B49 and B81 directly align with the vertical substructures at Z49 and Z81, respectively. The CO peak at B59 is also co-located with the Z56 feature in AS~209. Both changes in slope in IM~Lup and GM~Aur are spatially associated with CO features at B106 and B137, respectively. In IM~Lup, we also find that the D360 gap in line emission may be spatially related to the Z375 vertical dip. Given that both line emission peaks and gaps show some spatial association with surface substructures, this may point toward multiple mechanisms producing these surface features.

\section{Discussion} \label{sec:discussion}

\subsection{Comparison to previous results}

\citet{Pinte18} were the first to demonstrate an approach to directly extract CO emission surfaces in moderate resolution observations of IM~Lup. Similar methods were then used by \citet{Teague18_kinematic,Teague18_AS209, Teague19Natur} and \citet{Rich21} to constrain the emitting layers in AS~209 and HD~163296. Below, we compare these previous results to the high spatial resolution MAPS observations and comment on the relatively consistency and any salient differences.

\subsubsection{IM~Lup} \label{sec:im_lup_prev_results}

\citet{Pinte18} found CO and $^{13}$CO 2--1 emission heights of $z/r \approx 0.325$ and $z/r \approx 0.125$ in the inner flared disk region, i.e., $r<300$~au, using ${\sim}$0\farcs4 resolution observations. In this same region, we find a considerably more elevated CO surface of $z/r \gtrsim 0.5$ and $^{13}$CO of $z/r\approx0.2$. The differences in the derived heights are likely the result of our improved angular resolution, allowing for better separation of the front and back disk sides, e.g., see Figure \ref{fig:Band3_Surface}, which demonstrates the limitations of poorer spatial resolutions. \citet{Pinte18} also found that the CO 2--1 surface flattens out beyond 300~au, while our higher sensitivity observations reveal that this is due to a local minimum (Z375), and that globally the CO surface continues to rise out to 550~au due to diffuse large radii CO emission \citep[e.g.,][]{Cleeves16}. \citet{Pinte18} do not detect vertical substructures in any of their CO surfaces, but this is likely a consequence of their modest spatial resolution.

We also find consistent brightness temperature distributions to those of \citet{Pinte18}, but with overall systematically warmer temperatures by ${\sim}$5~K. This systematic offset is likely due to beam dilution in \citet{Pinte18} whose larger beam (${\sim}4\times$) would have smeared out some of this emission. Overall, these comparisons illustrate the importance of high spatial resolutions in accurately constraining emission surfaces. At more moderate spatial resolution, estimates of the heights of emitting layers and gas temperatures are both underestimated. A detailed exploration of the effect of spatial resolution on extracted surfaces within the MAPS disk is found in Appendix \ref{sec:app:spatial_res_on_surfaces}. In short, we find that surfaces derived from images with beam sizes between 0\farcs12 and 0\farcs2 are consistent, suggesting that the results presented here are not underestimated due to insufficient angular resolutions.

\citet{Rich21} fit the CO 2--1 emission surfaces in the IM~Lup disk using the DSHARP datacubes \citep{Andrews18} with a similar extraction method. These images have a spatial resolution of ${\sim}$0\farcs12, which is comparable to the MAPS resolution. Their CO 2--1 surface is nearly identical to the one we derived with the MAPS data and shows the same slope change at ${\sim}$110~au and a localized dip around 375~au.

\subsubsection{AS~209} \label{sec:as_209_prev_results}

\citet{Teague18_AS209} derived the CO 2--1 emission surface in the AS~209 disk using the same approach as \citet{Pinte18} using high spatial resolution (${\sim}$0\farcs2) CO line data. The authors found a surface with $z/r \sim 0.2$, which is the same $z/r$ we derived. However, they find a continually rising surface out to 300~au, but our surface begins to plateau and turnover at ${\gtrsim}$150~au. This difference may be due to their factor of two coarser spatial resolution. \citet{Teague18_AS209} also model the CO 2--1 emission surface that best reproduces the observed deviations in rotation velocities. Unlike the directly mapped surface, which is mostly smooth, the modeled surface has wave-like vertical substructures that appear very similar to those seen in the MAPS CO 2--1 emitting surface.

\subsubsection{HD~163296} \label{sec:hd_163296_prev_results}

\citet{Teague19Natur} mapped the CO 2--1 emission surface in the HD~163296 disk at an angular resolution of ${\sim}$0\farcs1. The authors constrained an emission surface out to a radius of 4$^{\prime \prime}$ that is nearly identical to one that we derive. In the same disk but using lower resolution observations (${\sim}$0\farcs25), \citet{Teague18_kinematic} modeled the emitting layer of C$^{18}$O 2--1 as a Gaussian process and found a typical $z/r \approx 0.1$, consistent with our C$^{18}$O surface. \citet{Teague18_kinematic} also found slight dips in their emission surface at the millimeter gap locations, i.e., ${\sim}$50, 80, 130~au (once rescaled to the \textit{Gaia} distance). The first two depressions correspond to the Z49 and Z81 dips in our surface, while the 130~au dip lies beyond the turnover region in our data. The discrepancy at large radii between these surfaces may be due in part to their larger beam size (about twice that of the MAPS beam) as well as a different approach to surface extraction. Nonetheless, they are still broadly consistent with one another.

\citet{Rich21} also used the DSHARP images (${\sim}$0\farcs1) to extract the CO 2--1 surface of HD~163296. The authors find a surface that is very consistent, i.e., $z/r\approx 0.35$, with that derived from the MAPS data. However, the authors do not resolve the inner dip at 46~au that is clearly seen in the MAPS surfaces.

\subsection{Origins of emission layer heights} \label{sec:orgins_of_layer_heights}

While the majority of emitting surfaces show a consistent general behavior, i.e., are well-described by an exponentially-tapered power law profile, there is considerable variation in their emitting heights. Here, we briefly explore possible mechanisms that may be important in setting the heights of disk emitting layers. Specifically, to explore the origins of the observed diversity in $z/r$ structure across disks, i.e., from $z/r \gtrsim 0.5$ to $z/r \lesssim0.1$, we searched for trends between physical parameters among the MAPS sources. While differences between isotopologues within disks are expected, differences between disks in the same isotopologues reveal variations in radiation fields and thermal, density, or CO abundance structures. To ensure a consistent comparison, we focus on the typical $z$/$r$ in the inner 150~au of each disk. We only consider the CO surfaces, as they are well-constrained in all disks and show a sufficiently wide range of $z$/$r$ values.

\begin{figure}[!ht]
\centering
\includegraphics[width=\linewidth]{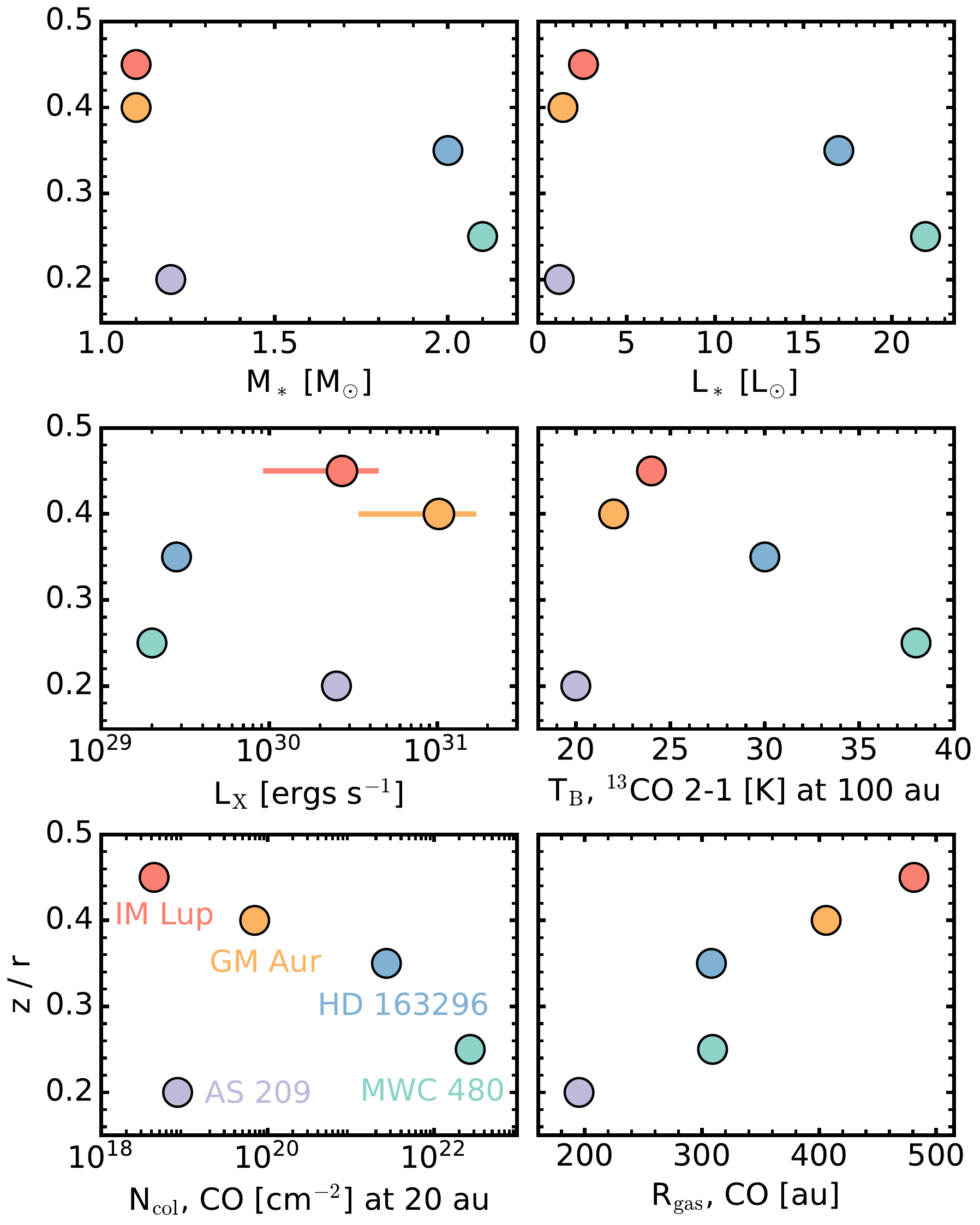}
\caption{Average $z/r$ of $^{12}$CO 2--1 emission heights versus disk physical properties for the MAPS sources. Stellar masses and bolometric luminosities, taken from \citet{oberg20} and references therein, while X-ray luminosities references are: IM~Lup \citep{Cleeves17}, AS~209 \citep{Walter81}, GM~Aur \citep{Espaillat19}, HD~163296 \citep{Gunther09}, and MWC~480 \citep{Grady10}. Horizontal bars indicate observed intrinsic X-ray variability. Brightness temperatures of $^{13}$CO 2--1 at 100~au are interpolated directly from the T$_{\rm{}B}$ radial profiles in Figure \ref{fig:Tb_CO_B6_profiles}, so they sometimes differ from the fitted $T_{100}$ in Table \ref{tab:radial_temperature_plaw_fits}. If we instead adopted the fitted $T_{100}$ values, the conclusions are unchanged. Peak CO column densities are from \citet{zhang20} (and see Figure \ref{fig:Ncol_vs_vertical_substr}), while the CO 2--1 disk sizes are from \citet{law20b}.}
\label{fig:Corr_figure}
\end{figure}

As irradiation from the central star plays a large role in setting the shape of the emitting layers \citep{Dullemond01, Dullemond04a, Dullemond04b}, we first consider whether differences in $z$/$r$ could be explain by differences in incident stellar radiation. In Figure \ref{fig:Corr_figure}, we identify a tentative negative trend between bolometric stellar luminosity and the $z/r$ of $^{12}$CO 2--1 emission surfaces as well as a modest positive association between X-ray luminosity and emission height. However, in both cases, AS~209 is an obvious outlier with an emission surface that is substantially flatter than the other two T Tauri sources IM~Lup and GM~Aur, which possess similar stellar and X-ray luminosities.

Another parameter that may set disk emitting layer heights is the temperature of the vertically isothermal layer \citep[e.g.,][]{Qi19}. To check this, we compared the $^{13}$CO 2--1 gas temperatures at 100~au with emission heights in Figure \ref{fig:Corr_figure}. With the exception of AS~209, we find a negative trend, where the warmer temperatures of the two disks around the Herbig~Ae stars have flatter surfaces, while the cooler temperatures of those around the T Tauri stars IM~Lup and GM~Aur have higher $z/r$ surfaces. We find a similar association if we instead consider the midplane temperature estimates derived from the thermo-chemical models of \citet{zhang20}.

We next consider the physical properties of the gas itself, i.e., total disk size and column densities. We find a tight positive trend between CO 2--1 disk size and the $z/r$ of CO surfaces, as shown in Figure \ref{fig:Corr_figure}. However, the surfaces of small disks turnover at smaller radii, which may affect the $z$/$r$, but this trend remains unchanged if we instead compare using the turnover radius for each disk. Assuming that H$_2$ number density scales with CO column densities, we expect more dust grains in the disk upper layers due to increased dynamical gas-grain coupling. This, may in turn, manifest as higher $z$/$r$ surfaces. However, if we compare peak CO column densities, i.e., $<50$~au, versus $z$/$r$, we find an inverse relation, with AS~209 being a notable outlier to this trend. Taken together, this suggests that larger disks with lower column densities may preferentially exhibit elevated emitting surfaces.

However, these source characteristics are not all independent, since the mass of the central star either sets or influences many of them. Therefore, we also compared the stellar mass and emission surface $z/r$ in Figure \ref{fig:Corr_figure} and found a negative trend, very similar to that of the stellar luminosity. The stellar mass sets the stellar luminosity, including the X-ray luminosity (with lower mass stars being more active), which in turn controls the disk temperature structure. Physically, warmer disks may be expected to result in increasingly flared surfaces, but this is the opposite of what we find in Figure \ref{fig:Corr_figure}. As vertical surfaces are set by the balance of pressure and gravity, disks around more massive stars should, in contrast, exhibit flatter surfaces. Since stellar mass positively correlates with disk mass\footnote{Literature M$_{*}$-M$_{\rm{disk}}$ correlations are typically derived in the optically thin limit, but as disk continuum emission may be partially optically thick, the estimated disk masses should be considered lower bounds \citep[e.g.,][]{Zhu19, Andrews20ARAA}.} \citep[e.g.,][]{Andrews13, Ansdell16, Pascucci16}, this scenario is consistent with the observed trends and suggests that stellar mass is the dominant factor in setting emission surface heights. Thus, the majority of observed trends may simply be tracing the impact of varying stellar masses and the effects on the surrounding disks.

Overall, however, we caution that this small and highly biased sample of disks limits generalized conclusions. A survey aimed at targeting CO lines in a large set of moderately inclined disks with sufficient resolution and sensitivity is needed to provide further constraints on the origins and distribution of the heights of disk emitting layers.

\subsection{Origins of Vertical Substructures} \label{sec:substr_discussion}

The emitting surfaces show several dips in vertical heights, which may have their origins in a variety of mechanisms. They may be due to CO depletion, i.e., decreased CO column density, decreases in total H$_2$ surface density but with constant CO abundance, or true geometrical features, e.g., warps. Here, we focus on the first two explanations and note that changes in CO abundance suggest a chemical origin, while decreases in H$_2$, hint at dynamical, planet-based origins. 

\begin{figure*}[ht]
\centering
\includegraphics[width=\linewidth]{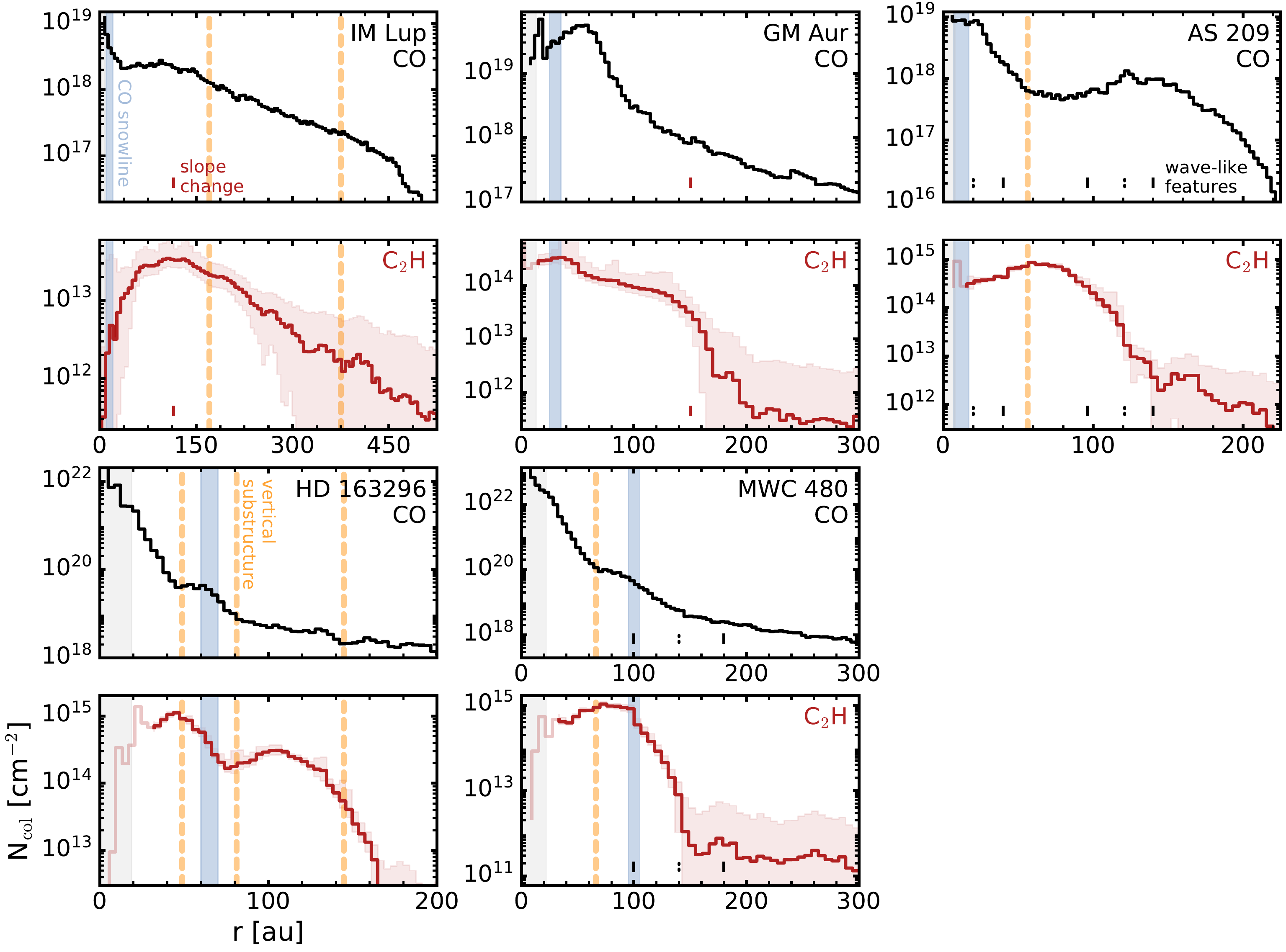}
\caption{Column density profiles for CO \citep{zhang20} and C$_2$H \citep{guzman20} for all MAPS sources versus locations of vertical substructures in CO, $^{13}$CO, and C$^{18}$O 2--1 emission lines. Orange dashed lines indicates vertical dips listed in Table \ref{tab:vertical_substructures}. The CO snowlines from \citet{zhang20} are shaded in blue. Red solid lines indicate the radial locations of changes in emission surface slope. The low-amplitude, wave-like features in AS~209 and MWC~480 are shown in black with dashed and solid lines marking troughs and peaks, respectively.}
\label{fig:Ncol_vs_vertical_substr}
\end{figure*}

The chemical explanation requires CO gas to be sufficiently depleted at the locations of vertical substructures. Alignments of vertical substructures with CO column density gaps, while suggestive, are not conclusive proof of a chemical origin, as gas surface density perturbations and chemical processing are often degenerate in models \citep[e.g.,][]{alarcon20}. However, if these features have their origins in chemistry, the depletion of gas-phase CO should lead to higher C/O ratios, causing an increase in the column density of molecules such as C$_2$H \citep{Bergin16, alarcon20}. To test this possibility, we compare the column density profiles of CO \citep{zhang20} and C$_2$H \citep{guzman20} with the identified vertical substructures in Figure \ref{fig:Ncol_vs_vertical_substr}. The vertical substructures Z56 in AS~209, Z46 in HD~163296, and Z66 in MWC~480 are all associated with CO column density depletions and C$_2$H enhancements. In contrast, Z81 and Z145 in HD~163296, as well as Z170 and Z375 in IM~Lup are not. This suggests that chemical conversion of CO into other species may provide at best a partial explanation for the observed vertical substructures.

The dynamical explanation, i.e., if vertical substructures are caused by forming-planets, instead requires drops in total H$_2$ gas surface density. Perhaps, in this case, we expect vertical substructures to be associated with gaps in the millimeter-sized grains. As large grains should be concentrated in gas pressure maxima, this means that dust gaps will correspond to pressure minima and be associated with drops in the total H$_2$ gas surface density. As discussed in Section \ref{sec:mm_continuum_vs_chemical_substr}, all of the MAPS disks show some degree of spatial association between millimeter continuum gaps and vertical substructures. Below, we consider the plausibility of this interpretation for each MAPS disk:

In AS~209, the Z56 substructure is radially coincident with a deep gap in H$_2$ \citep{Teague18_AS209}. Moreover, \citet{Fedele18} showed that either a single Saturn mass planet at 95~au, or a second ($<0.1~M_{\rm{Jup})}$ planet 57~au, reproduced the observed continuum profile. Thus, a planetary-origin for Z56 (and the larger wave-like structure) in AS~209 is possible, but c.f., \citet{alarcon20}, who constrain the mass of a putative planet at 100~au to be ${<}0.2$~M$_{\rm{Jup}}$.

In HD~163296, the derived CO surfaces, and in particular, the measured depths of vertical substructures, in HD~163296 most closely match the models of \citet{Rab20} that include deep gas gaps, i.e., similar depletion as for the dust, at the locations of the observed millimeter continuum gaps. Similarly, the gas gap models (versus that of CO depletion) from \citet{calahan20b} are better able to reproduce the Z81 dip in the C$^{18}$O emission surface. As HD~163296 is also believed to host three Jupiter-mass planets \citep{Pinte18, Teague18_kinematic}, this offers a plausible explanation for the vertical substructures in this disk.

In IM~Lup, GM~Aur, and MWC~480, the plausibility of vertical substructures having their origins in planets is less clear. In IM~Lup, \citet{Pinte20} reported a tentative localized deviation from Keplerian rotation at 117~au, which is thought to be due to a planetary perturber. Intriguingly, this is at the same radius where we observe a change in the slope of the CO 2--1 emitting surface. GM~Aur has been suggested to have a 0.1--0.4~M$_{\rm{Jup}}$ planet at 67~au based on the width of the nearby dust gap \citep{Huang20_GMAur}, but no vertical substructures are observed near this radius. The wave-like feature in MWC~480 shows broad associations with dust gaps, and \citet{teague20} propose a planet at 245~au, which is driving the wave-like perturbations. 

Regardless of the specific mechanisms responsible for these vertical substructures, the MAPS data suggest that emitting surfaces are far from smooth. The locations, depths, and widths of such features provide important inputs to disk thermo-chemical models and serve as powerful probes of the planet formation. As such, they may also offer another promising mechanism to infer the existence of embedded, newly-forming planets in disks.

\section{Conclusions} \label{sec:conclusions}

We present a detailed analysis of the vertical distribution of molecules and their emitting surfaces in high angular resolution observations in five protoplanetary disks from MAPS. We conclude the following:

\begin{enumerate}
  \item CO emission traces the most elevated regions $z/r > 0.3$, while the less abundant $^{13}$CO and C$^{18}$O probe deeper into the disk $z/r \approx 0.1$--$0.2$. These heights correspond to approximately 3 and 1 scale heights, respectively. 
  \item In the disks around the T~Tauri star AS~209 and Herbig~Ae stars HD~163296 and MWC~480, C$_2$H and HCN emission heights are also measurable and they emit from $z/r \lesssim 0.1$, a region relatively close to the planet-forming disk midplane.
  \item The NIR surfaces, which trace micron-sized dust, of HD~163296 and IM~Lup, are lower than the CO 2--1 emission surface and lie at or slightly above that of $^{13}$CO 2--1.
  \item We derive radial temperature distributions for all CO isotopologues and use them to estimate full 2D, ($r$, $z$) empirical temperature models for each disk.
  \item Emission surfaces present substructures in the form of vertical dips, often seen in more than one CO isotopologue, and are detected in a majority of MAPS disks. 
  \item The wide range of vertical emission heights across the sample indicates a diversity in thermal, density, or CO abundance structures. Tentative trends suggest that star+disk systems with lower stellar masses and luminosities, as well as larger CO disk sizes exhibit the most elevated CO line-emitting surfaces. However, a larger sample of disks with well-constrained disk emitting layers is required to better understand what sets emitting layer heights in disks.   
  \item At least some, and possibly the majority, of vertical disk substructures have their origins in local H$_2$ surface density drops due to embedded planets. Others may have their origin in chemical effects, namely local reductions in CO abundance and thus CO optical depth.
\end{enumerate}

Overall, we have shown an effective method for extracting the emitting layers for a sample of disks and emission lines. As disks are highly structured both radially and vertically, emission surfaces in a set of lines with varying optical depths, e.g., CO isotopologues, provide direct observational constraints on the overall 2D disk structure. Moreover, these surfaces serve as critical inputs to thermo-chemical models of disks, which are necessary to not only understand the true origins of vertical gas structures but also to connect observed molecular emission to midplane abundances, and therefore the chemical environment within which planets form.

\section{Value-Added Data Products} \label{sec:VADPs}

The MAPS Value-Added Data Products described in this work can be accessed through the ALMA Archive via \url{https://almascience.nrao.edu/alma-data/lp/maps}. An interactive browser for this repository is also available on the MAPS project homepage at \url{http://www.alma-maps.info}.

For each combination of data processing (individual measurements, radially-binned, and moving average), the following data products are available:

\begin{itemize}
    \item Emission surfaces
    \item Gas temperature structures, radial and full 2D ($r$, $z$) profiles
    \item Python script to generate the data products
\end{itemize}

Each of these VADPs are provided for CO 2--1, $^{13}$CO 2--1, and when available, C$^{18}$O 2--1 in all MAPS disks, and for C$_2$H 3--2 and HCN 3--2 in HD~163296. The naming scheme for these VADPs is as follows: [disk]\_[line]\_[frequency]\_[resolution]\_[datatype], where datatype is: ``individual measurements," ``radially-binned," or ``moving average." Additional data products associated with the MAPS Large Program, including line image cubes \citep[see Section 9,][]{czekala20} and radial profiles and moment maps \citep[see Section 7,][]{law20b}, are also available.

\acknowledgments
The authors thank the anonymous referee for valuable comments that improved both the content and presentation of this work. This paper makes use of the following ALMA data: ADS/JAO.ALMA\#2018.1.01055.L. ALMA is a partnership of ESO (representing its member states), NSF (USA) and NINS (Japan), together with NRC (Canada), MOST and ASIAA (Taiwan), and KASI (Republic of Korea), in cooperation with the Republic of Chile. The Joint ALMA Observatory is operated by ESO, AUI/NRAO and NAOJ. The National Radio Astronomy Observatory is a facility of the National Science Foundation operated under cooperative agreement by Associated Universities, Inc.

C.J.L. acknowledges funding from the National Science Foundation Graduate Research Fellowship under Grant DGE1745303. R.T. and F.L. acknowledge support from the Smithsonian Institution as a Submillimeter Array (SMA) Fellow. J.B. acknowledges support by NASA through the NASA Hubble Fellowship grant \#HST-HF2-51427.001-A awarded  by  the  Space  Telescope  Science  Institute,  which  is  operated  by  the  Association  of  Universities  for  Research  in  Astronomy, Incorporated, under NASA contract NAS5-26555. K.I.\"O. acknowledges support from the Simons Foundation (SCOL \#321183) and an NSF AAG Grant (\#1907653). I.C. was supported by NASA through the NASA Hubble Fellowship grant \#HST-HF2-51405.001-A awarded by the Space Telescope Science Institute, which is operated by the Association of Universities for Research in Astronomy, Inc., for NASA, under contract NAS5-26555. S.M.A. and J.H. acknowledge funding support from the National Aeronautics and Space Administration under Grant No. 17-XRP17 2-0012 issued through the Exoplanets Research Program. J.H. acknowledges support for this work provided by NASA through the NASA Hubble Fellowship grant \#HST-HF2-51460.001-A awarded by the Space Telescope Science Institute, which is operated by the Association of Universities for Research in Astronomy, Inc., for NASA, under contract NAS5-26555. Y.A. acknowledges support by NAOJ ALMA Scientific Research Grant Code 2019-13B, and Grant-in-Aid for Scientific Research 18H05222 and 20H05847. E.A.B., A.D.B., and F.A. acknowledge support from NSF AAG Grant (\#1907653). A.S.B. acknowledges the studentship funded by the Science and Technology Facilities Council of the United Kingdom (STFC). J.B.B. acknowledges support from NASA through the NASA Hubble Fellowship grant \#HST-HF2-51429.001-A, awarded by the Space Telescope Science Institute, which is operated by the Association of Universities for Research in Astronomy, Inc., for NASA, under contract NAS5-26555. J.K.C. acknowledges support from the National Science Foundation Graduate Research Fellowship under Grant No. DGE 1256260 and the National Aeronautics and Space Administration FINESST grant, under Grant no. 80NSSC19K1534. G.C. is supported by NAOJ ALMA Scientific Research Grant Code 2019-13B. L.I.C. gratefully acknowledges support from the David and Lucille Packard Foundation and Johnson \& Johnson's WiSTEM2D Program. V.V.G. acknowledges support from FONDECYT Iniciaci\'on 11180904 and ANID project Basal AFB-170002. J.D.I. acknowledges support from the Science and Technology Facilities Council of the United Kingdom (STFC) under ST/T000287/1. R.L.G. acknowledges support from a CNES fellowship grant. Y.L. acknowledges the financial support by the Natural Science Foundation of China (Grant No. 11973090). F.M. acknowledges support from ANR of France under contract ANR-16-CE31-0013 (Planet-Forming-Disks) and ANR-15-IDEX-02 (through CDP ``Origins of Life"). H.N. acknowledges support by NAOJ ALMA Scientific Research Grant Code 2018-10B and Grant-in-Aid for Scientific Research 18H05441. L.M.P. acknowledges support from ANID project Basal AFB-170002 and from ANID FONDECYT Iniciaci\'on project \#11181068. K.R.S. acknowledges the support of NASA through Hubble Fellowship Program grant HST-HF2-51419.001, awarded by the Space Telescope Science Institute, which is operated by the Association of Universities for Research in Astronomy, Inc., for NASA, under contract NAS5-26555. T.T. is supported by JSPS KAKENHI Grant Numbers JP17K14244 and JP20K04017. Y.Y. is supported by IGPEES, WINGS Program, the University of Tokyo. M.L.R.H. acknowledges support from the Michigan Society of Fellows. C.W. acknowledges financial support from the University of Leeds, STFC and UKRI (grant numbers ST/R000549/1, ST/T000287/1, MR/T040726/1). K.Z. acknowledges the support of the Office of the Vice Chancellor for Research and Graduate Education at the University of Wisconsin – Madison with funding from the Wisconsin Alumni Research Foundation, and support of the support of NASA through Hubble Fellowship grant HST-HF2-51401.001. awarded by the Space Telescope Science Institute, which is operated by the Association of Universities for Research in Astronomy, Inc., for NASA, under contract NAS5-26555.

%

\vspace{5mm}
\facilities{ALMA}


\software{Astropy \citep{astropy_2013,astropy_2018}, \texttt{bettermoments} \citep{Teague18_bettermoments}, CASA \citep{McMullin_etal_2007}, \texttt{disksurf} (\url{https://github.com/richteague/disksurf}), \texttt{emcee} \citep{Foreman_Mackey13}, \texttt{GoFish} \citep{Teague19JOSS}, Matplotlib \citep{Hunter07}, NumPy \citep{vanderWalt_etal_2011}, SciPy \citep{Virtanen_etal_2020}} \clearpage



\appendix

\section{Surface Filtering and Extraction} \label{sec:app:surface_extraction}

\noindent In this section, we describe in detail the functionality of \texttt{disksurf} and how we used it to extract emission surfaces from line image cubes:

Before extracting surfaces, we applied the following two data filtering steps. First, a radially-varying clip was used to remove pixels that are not related to the peak of the line. This was done by calculating an azimuthally-averaged profile of the peak surface brightness and then clipping values that are more than 1$\sigma$ away. This clip threshold was increased to 2$\sigma$ for lower SNR lines, e.g., C$_2$H 3--2 and HCN 3--2. Then, we performed a 1D smoothing\footnote{We note that this smoothing is performed prior to and as a part of the extraction process and is thus distinct from the radial binning of the extracted data used subsequently to increase the SNR.} to better define the peaks using a Gaussian kernel with a full-width at half-maximum (FWHM) equal to half of the beam major axis FWHM. These two steps were found to significantly improve the ability of the code to identify the emission peaks and minimize contamination from background thermal noise.

We used the \texttt{get\_emission\_surface} function to extract the deprojected radius $r$, emission height $z$, surface brightness I$_{\nu}$, and channel velocity $v$ for each pixel associated with the emitting surface. This requires knowledge of the disk inclination and position angle in order to correctly account for the deprojection. We adopted these values from Table 1 in \citet{oberg20} with disk inclinations ranging from 37.0$^{\circ}$ (AS~209) to 53.2$^{\circ}$ (GM~Aur).

\begin{figure*}[!ht]
\centering
\includegraphics[width=\linewidth]{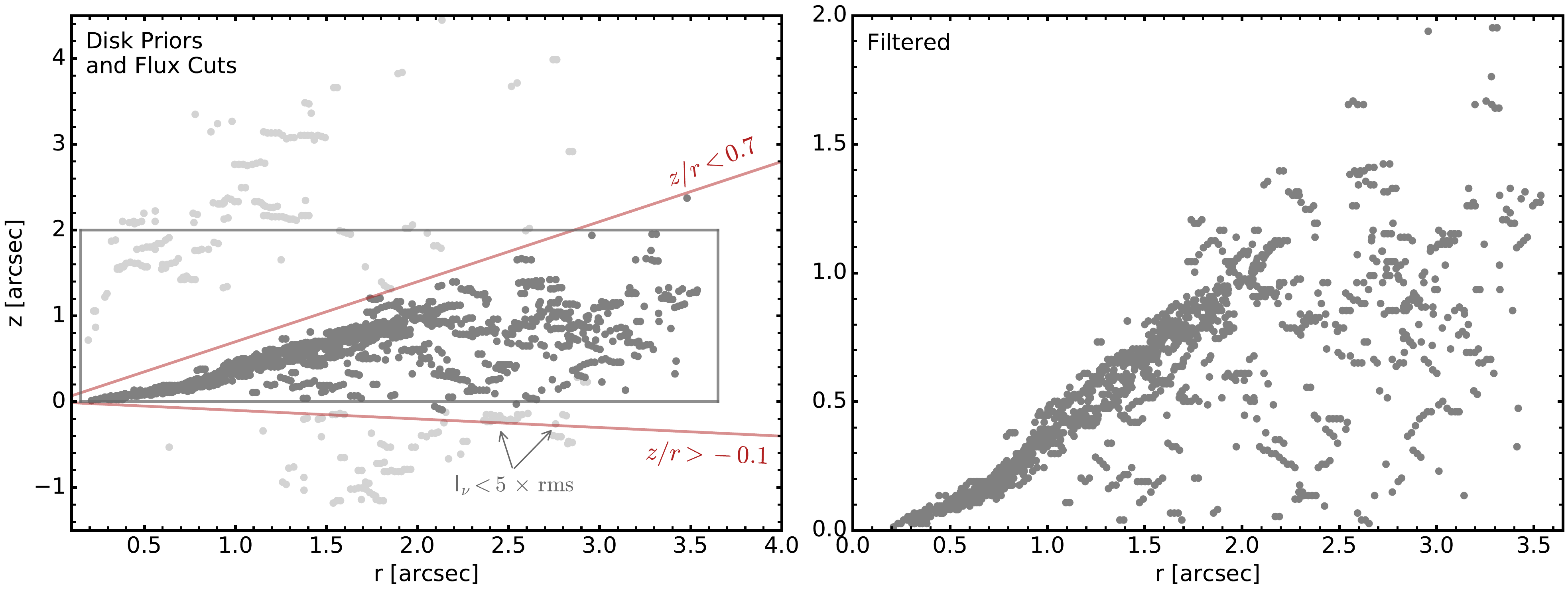}
\caption{Clipping based on disk priors and flux (left) and filtered emission surfaces (right) for CO 2--1 in IM~Lup. The gray box in the left panel shows the field of view in the right panel. Light gray points are filtered according to the steps described in detail in Appendix \ref{sec:app:surface_extraction}.}
\label{fig:Example_filtering_appendix}
\end{figure*}

We then applied additional clipping based on our priors of disk physical structure. In particular, we removed extremely high $z/r$ values\footnote{This threshold was initially set to clip vertical heights exceeding $z/r=0.5$ for all disks but had to be increased to $z/r=0.7$ for CO 2--1 in IM~Lup and GM~Aur, due their very elevated surfaces.} and large negative $z$ values, as the emission must arise from at least the midplane. We allowed points with a small negative $z$ value, i.e.,  $z/r>-$0.1, to remain to avoid positively biasing our averages to non-zero $z$ values. We also filtered those points with low surface brightness (less than 5 times the image cube RMS) to ensure that noise did not significantly bias the derived surfaces. However, due to the lower SNR of the C$_2$H and HCN lines, we did not perform this clipping in any disk except for HD~163296. Figure \ref{fig:Example_filtering_appendix} shows an example of this process for CO 2--1 in IM~Lup.

\section{Full list of isovelocity contours} \label{sec:app:isovelocity}

\begin{figure*}[!ht]
\centering
\includegraphics[width=\linewidth]{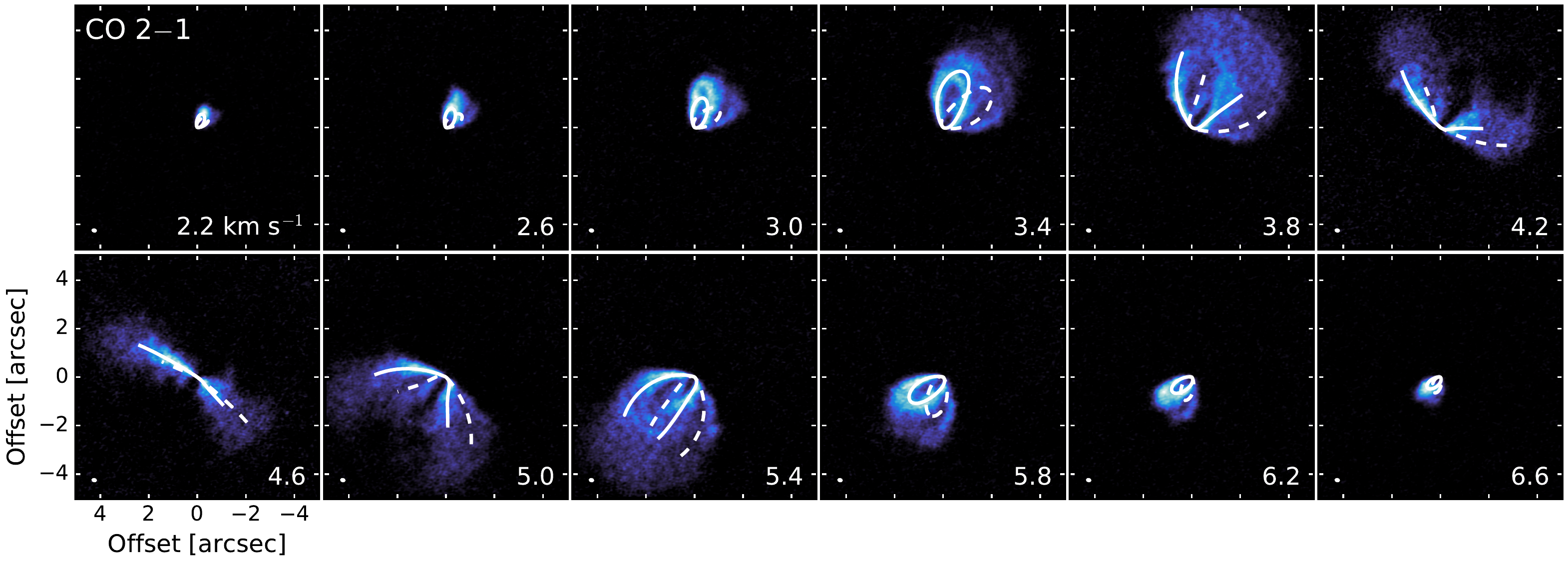}
\includegraphics[width=\linewidth]{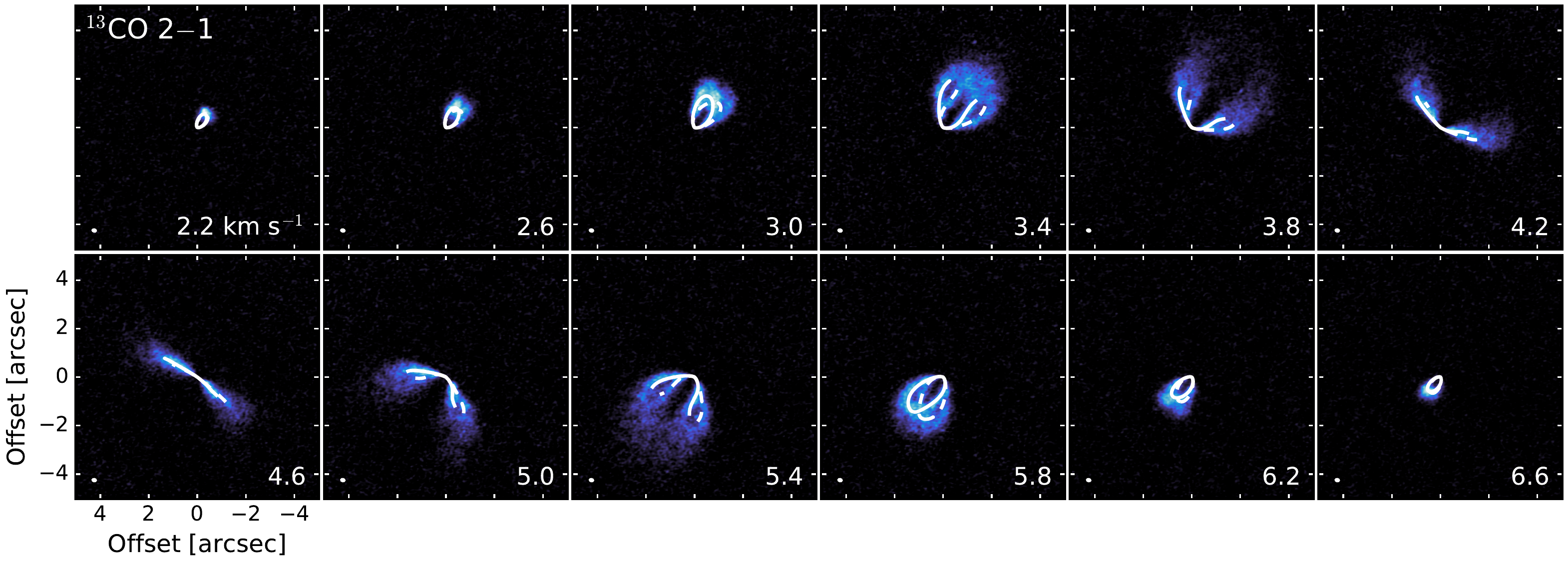}
\caption{Isovelocity contours of the best-fitting model, as indicated in Table \ref{tab:emission_surf}, for the CO 2--1 (top) and $^{13}$CO 2--1 (bottom) emitting surfaces in IM~Lup plotted for selected channels of the observed emission. Solid curves indicate the upper surface of the disk and dashed curves mark the lower surface. LSRK velocities are noted in the lower right corner. The synthesized beam is shown in the bottom left corner of each panel.}
\label{fig:Isovelocity_IM_Lup}
\end{figure*}

Isovelocity contours for the CO isotopologues are shown for the IM~Lup disk in Figure \ref{fig:Isovelocity_IM_Lup}. A full set of isovelocity contours for all MAPS disks are shown in Figure Set 1, which is available in the electronic edition of the journal. All isovelocity contours are calculated using updated dynamical masses taken from \citet{oberg20}, which are based on CO rotation profiles \citep{teague20}.

\section{Excitation and Band 3 CO Surfaces} \label{sec:app:Band3_excitation}
Since we also had access to $^{13}$CO 1--0, we were able to compare against the $^{13}$CO 2--1 line to see if we could identify any excitation-related effects in the emission surfaces, i.e., differing heights \citep[e.g.,][]{vanZadelhoff01, Dartois_ea_2003}. Due to the coarser spatial resolution and lower SNR of the 1--0 line, we did not attempt to extract the emission surfaces directly. Instead, we compared the $^{13}$CO 2--1 isovelocity contours derived from the parametric fit in Table \ref{tab:emission_surf} with the spatial distribution of the 1--0 line. To ensure a consistent comparison, we also included the tapered (0\farcs30) resolution $^{13}$CO 2--1 images. We checked C$^{18}$O 1--0, which was also covered by the MAPS observations, but it did not possess sufficient SNR for this comparison, so we instead focused on $^{13}$CO 1--0. At this lower resolution, only GM~Aur, HD~163296, and IM~Lup had sufficiently elevated $^{13}$CO 2--1 surfaces to allow for a meaningful comparison.

\begin{figure}[ht]
\centering
\includegraphics[width=0.6\linewidth]{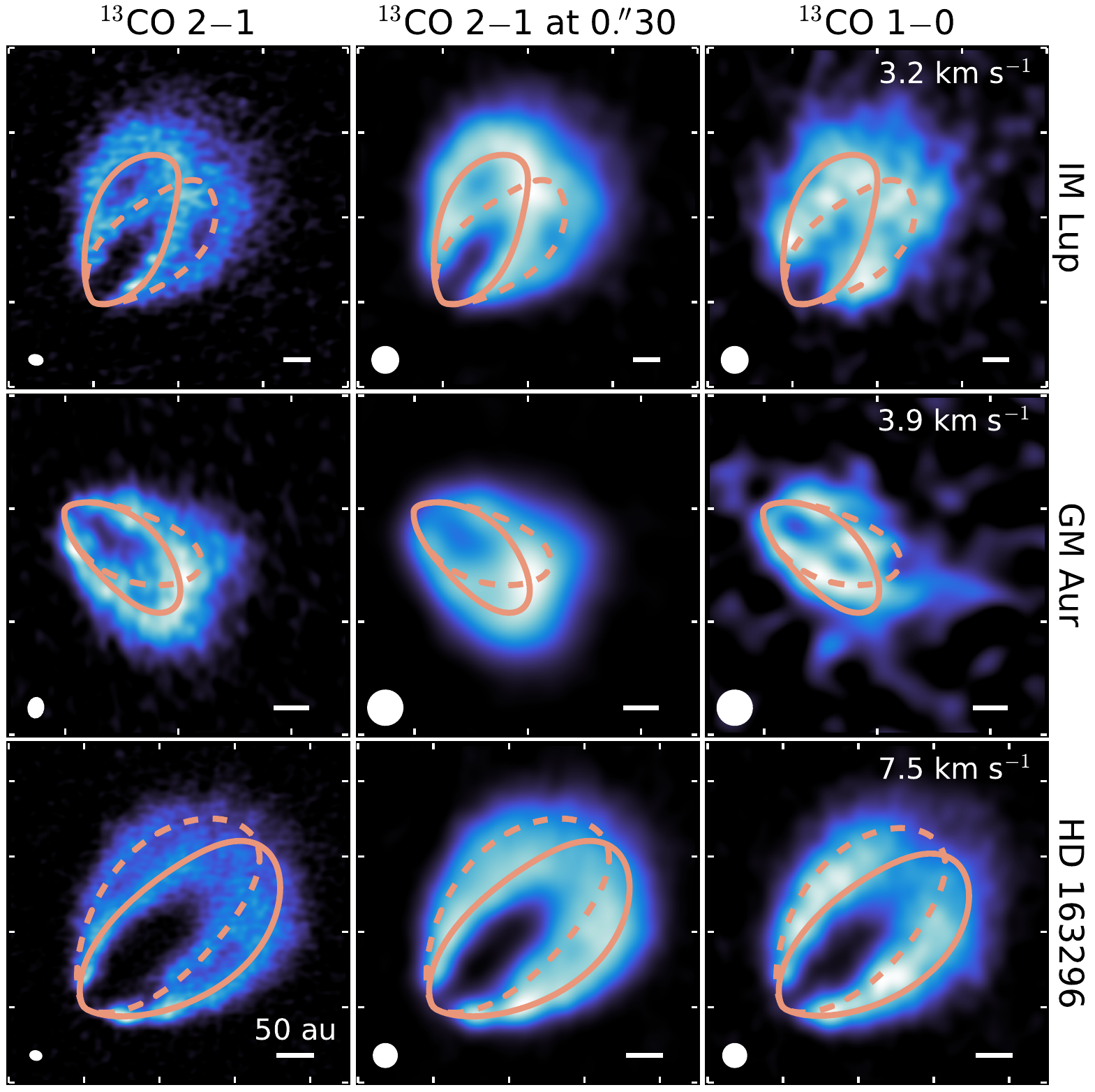}
\caption{Representative channels for IM~Lup, GM~Aur, and HD~163296 for the $^{13}$CO 2--1 full resolution (left column), $^{13}$CO 2--1, tapered to 0\farcs30 (middle column), and $^{13}$CO 1--0 (right column) images. The $^{13}$CO 2--1 isovelocity contours derived using the parametric fit in Table \ref{tab:emission_surf} are shown in pink. Solid curves indicate the upper surface of the disk and dashed curves mark the lower surface. LSRK velocities are noted in the upper right corner. The synthesized beam and a scale bar indicating 50~au is shown in the lower left and right corner, respectively, of each panel.}
\label{fig:Band3_Surface}
\end{figure}

Figure \ref{fig:Band3_Surface} shows isovelocity contours overlaid on a representative channel of $^{13}$CO 1--0 emission that should best show the emitting layers, if resolved. In IM~Lup and GM~Aur, line emission tracing the back side of the disks is visible in the tapered 2--1 image, but we cannot determine if the contours are consistent with low SNR emission surfaces in $^{13}$CO 1--0, or if the 1--0 line is truly flatter than the 2--1 line. In the case of HD~163296, the spatial resolution is insufficient to reveal any vertical disk structure in either the $^{13}$CO 2--1 tapered or 1--0 images, and we therefore cannot compare 2-1 and 1-0 emission layer heights. Thus, in order to infer the emitting heights of $^{13}$CO 1--0, we likely require both higher spatial resolution and SNR.

\section{Effects of spatial resolution on derived emission surfaces} \label{sec:app:spatial_res_on_surfaces}

\begin{figure}[ht]
\centering
\includegraphics[width=\linewidth]{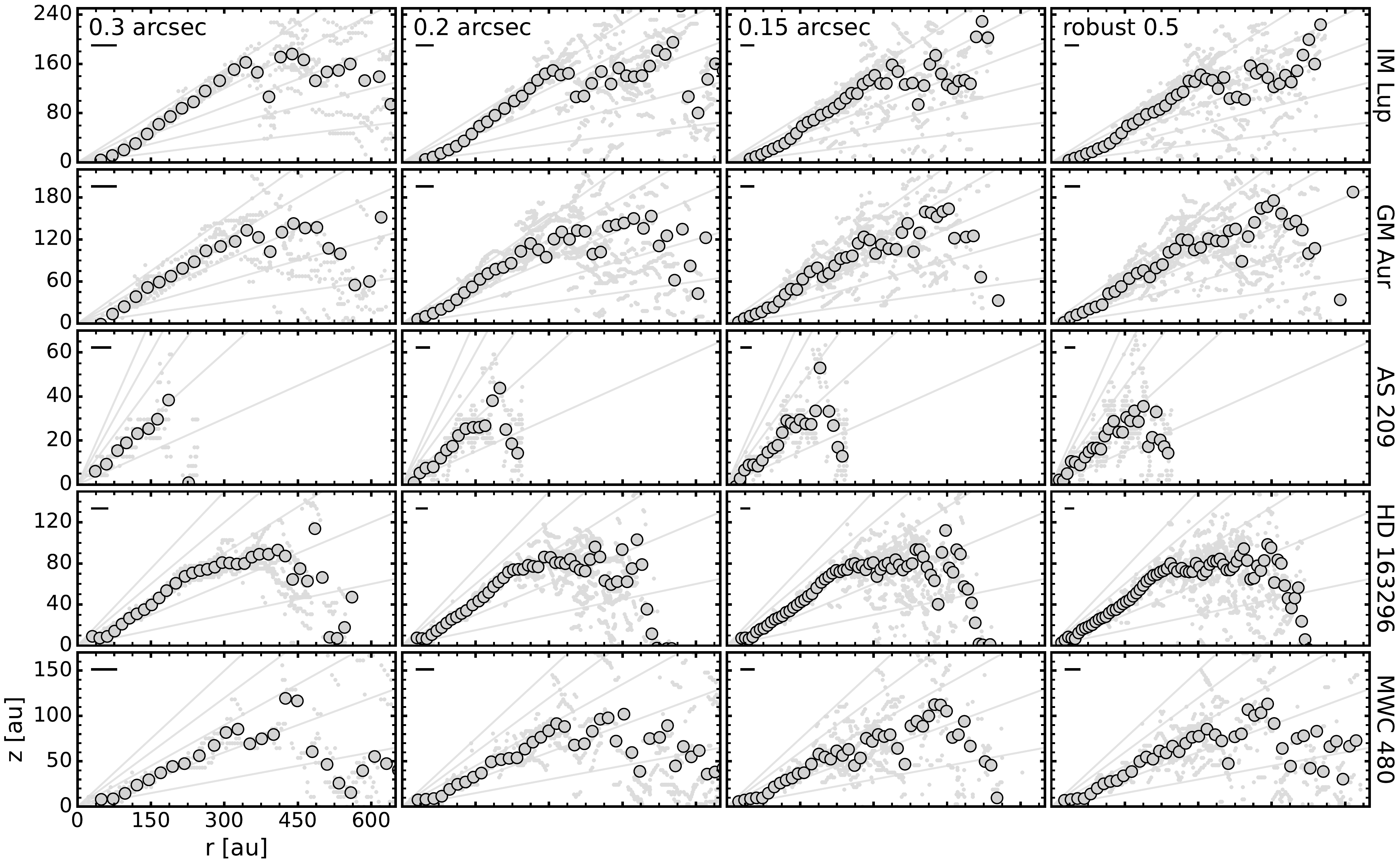}
\caption{$^{12}$CO 2--1 emission surfaces derived using images with different angular resolutions (see Table 5 in \citealt{oberg20}). Large, gray points show a consistent radial binning of 1/2$\times$ the beam major axis of each image, while small, light gray points represent individual measurements. Lines of constant $z/r$ from 0.1 to 0.5 in increments of 0.1 are shown in gray. The FWHM of the major axis of the synthesized beam is shown in the upper left corner of each panel.}
\label{fig:12CO_surfaces_v_resolution}
\end{figure}

\begin{figure}[ht]
\centering
\includegraphics[width=\linewidth]{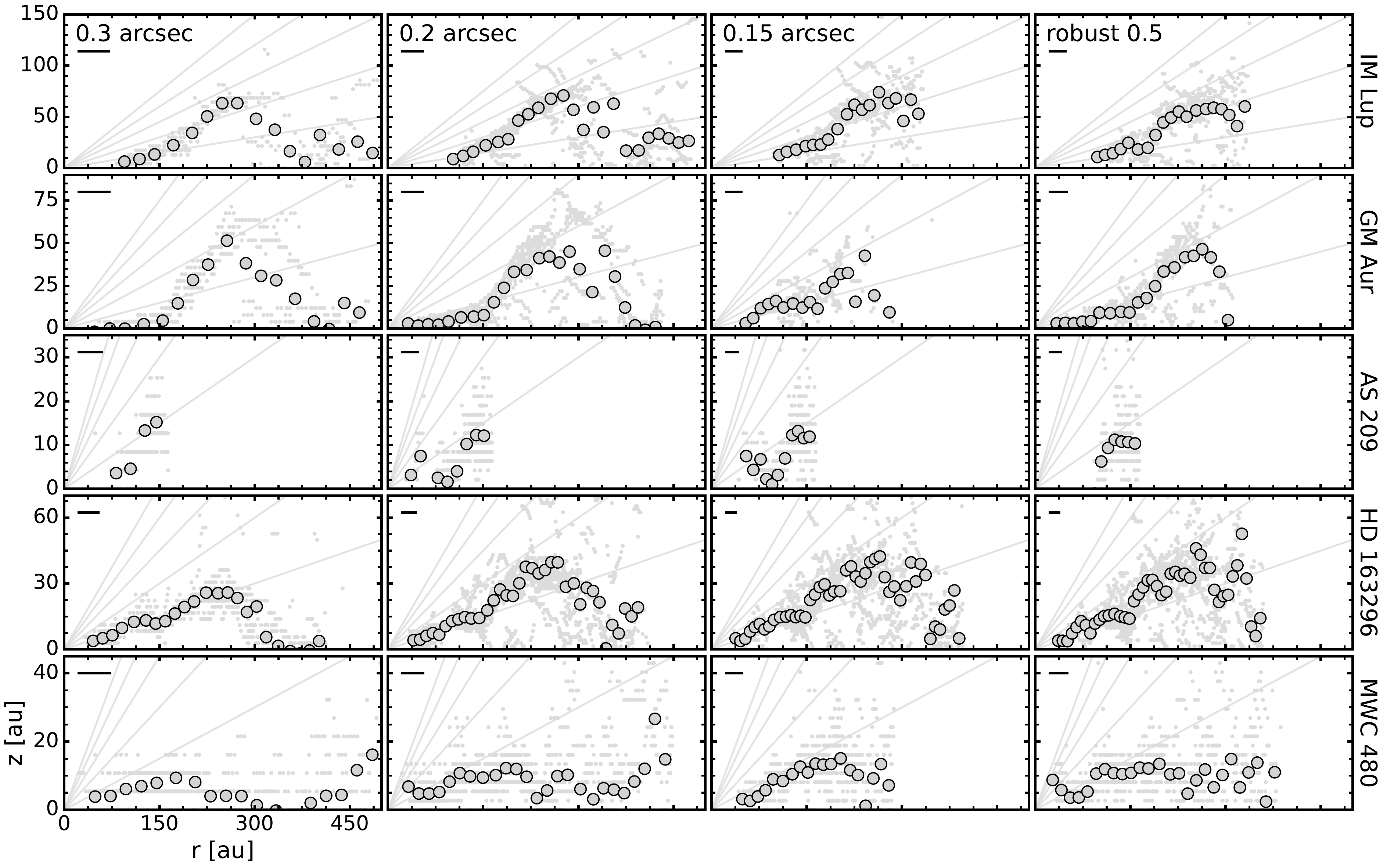}
\caption{$^{13}$CO 2--1 emission surfaces derived using different spatial resolutions. Otherwise, as in Figure \ref{fig:12CO_surfaces_v_resolution}.}
\label{fig:13CO_surfaces_v_resolution}
\end{figure}

\begin{figure}[ht]
\centering
\includegraphics[width=\linewidth]{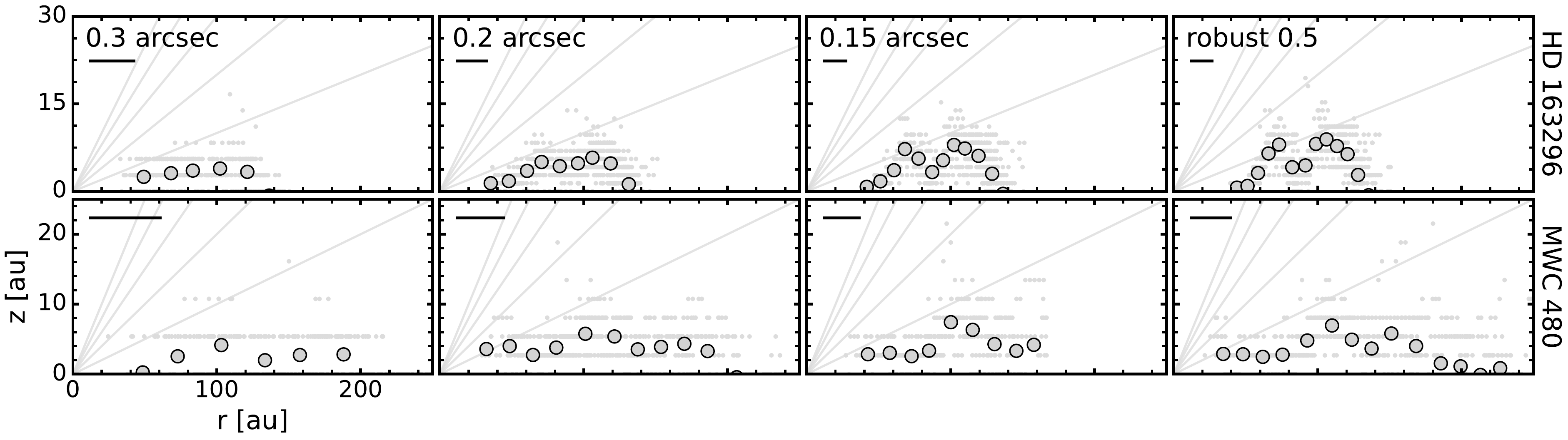}
\caption{C$^{18}$O 2--1 emission surfaces derived using different spatial resolutions for HD~163296 and MWC~480. The C$^{18}$O 2--1 surface in GM~Aur is not shown as it was only able to be extracted from the images generated with a robust parameter of 0.5, as only this image possessed a sufficiently high spatial resolution necessary to separate the front and back disk surfaces. Otherwise, as in Figure \ref{fig:12CO_surfaces_v_resolution}.}
\label{fig:C18O_surfaces_v_resolution}
\end{figure}

To extract emission surfaces using \texttt{disksurf}, the imagecubes must have some minimum angular resolution, i.e., the front and back sides of the disk must be sufficiently spatially resolved to be separable. This means that emission surfaces will be sensitive to the spatial resolution of the images used to derive them. To investigate the effects of spatial resolution on our surfaces, we repeated the surface extraction for CO, $^{13}$CO, and C$^{18}$O 2--1 using all angular resolutions, i.e., 0\farcs3, 0\farcs2, 0\farcs15, imaged as part of MAPS \citep{oberg20}. We then compared them to the surfaces derived from the images generated with a robust parameter of 0.5 used throughout this work. Figures \ref{fig:12CO_surfaces_v_resolution}, \ref{fig:13CO_surfaces_v_resolution}, and \ref{fig:C18O_surfaces_v_resolution} show the resulting surfaces.

The CO isotopologue surfaces are generally consistent across differing spatial resolutions. The lower resolutions (0\farcs3, 0\farcs2) occasionally underestimate the average $z/r$ surface height, e.g., $^{13}$CO and C$^{18}$O in HD~163296, and this effect is more conspicuous in intrinsically flatter surfaces, e.g., $^{13}$CO and C$^{18}$O in MWC~480 or those from rarer isotopologues. In highly-elevated surfaces like $^{12}$CO 2--1, the emission structure along a given column of pixels will be two well-separated Gaussians (and another two Gaussians for the backside of the disk). At lower spatial resolutions, these Gaussians are broadened, but do not overlap. Conversely, for $^{13}$CO or other emission lines with lower intrinsic $z/r$, these two components may overlap and thus lead to a single-peaked Gaussian with a $z/r$ that approaches 0. Even for those highly flared surfaces, images with higher spatial resolutions are preferable, as they allow for the detection and characterization of vertical substructures, such as the dips, wave-like features, and slope changes seen in many of the MAPS disks (see Section \ref{sec:substructures_emisssion_surfaces}). For instance, the $^{12}$CO surface in HD~163296 appears nearly identical between the 0\farcs3 and robust=0.5 images, with the important exception of the Z46 dip, which can only identified in the 0\farcs15 and robust=0.5 images.

\newpage
\clearpage



\bibliography{MAPS2-Vertical}{}
\bibliographystyle{aasjournal}



\end{document}